\documentclass[aps,showpacs,nofootinbib]{revtex4}
\usepackage{amsmath} 
\usepackage{graphicx} 
\usepackage{color} 
\usepackage{epstopdf}

\begin{document}

\newcommand{\blue}[1]{\textcolor{blue}{#1}}
\newcommand{\new}{\blue}
\newcommand{\green}[1]{\textcolor{green}{#1}}
\newcommand{\modif}{\green}
\newcommand{\red}[1]{\textcolor{red}{#1}}
\newcommand{\attention}{\red}


\title{Hollowness Effect and Entropy in High Energy Elastic Scattering}

\author{S. D. Campos} \email{sergiodc@ufscar.br}
\affiliation{Departamento de F\'isica, Qu\'imica e Matem\'atica, Universidade Federal de S\~ao Carlos
Sorocaba, S\~ao Paulo CEP 18052780, Brazil}

\author{V. A. Okorokov} \email{VAOkorokov@mephi.ru; Okorokov@bnl.gov}
\affiliation{National Research Nuclear University MEPhI (Moscow
Engineering Physics Institute), Kashirskoe highway 31, 115409
Moscow, Russia}

\date{\today}

\begin{abstract}
This paper presents a qualitative explanation for the hollowness
effect based on the inelastic overlap function, claiming this
result is a consequence of fundamental thermodynamic processes.
Using the Tsallis entropy, one identifies the entropic index $w$
with the ratio of the collision energy to critical one in the
total cross-section. The integrated probability density function
is replaced by the inelastic overlap function, which
represents the probability of occurrence of an inelastic event depending on
both the collision energy and impact parameter. The Coulomb
potential, as well as the confinement potential, are used as naive
approaches to describe the (internal) energy of the colliding
hadrons. The Coulomb potential in the impact parameter picture is
not able to furnish any reliable physical result near the forward
direction. However, the confinement potential in the impact
parameter space results in the hollowness effect shown by the
inelastic overlap function near the forward direction.

\end{abstract}

\pacs{13.85.Dz;13.85.Lg}

\maketitle

\section{Introduction}\label{sec:1}

The proton-proton ($pp$) and antiproton-proton ($\bar{p}p$)
elastic scattering at high energies remain as one of the most
surprising open issues of the collision processes. Some of these
open questions may be solved by the new phase of the so-called
High Luminosity Large Hadron Collider (HL--LHC), which will
deliver 3 ab$^{-1}$ of integrated luminosity
\cite{CERN-YR-2019-007}. In the future, other issues certainly
will arise, resulting in the construction of novel models or the
improvement of the present-day ones.

As well-known, the geometric point of view of $pp$ and $\bar{p}p$
scattering is an important tool to describe its dynamics,
furnishing insights as well as approaches for the non-perturbative
QCD. Of course, (anti)protons are not point-like
objects, but in wide energy range, they behave
similar to the black disk picture \cite{anisovich_nikonov,V.V.Anisovich.M.A.Matveev.V.A.Nikonov.Int.J.Mod.Phys.A31.1645019.2016} of classical optics, as can be
observed in the experimental results obtained over last decades \cite{sdcvaocvm}.
Correspondence to the black disk means a large probability of absorption (approaches to unity, strictly speaking) for some range of the quantity used in a certain model. Usually, that quantity is the impact parameter $b$. It should be clarified that the above statement, concerning the behavior of the (anti)protons, describes only a qualitative approach to the black disk regime at an accuracy level $\sim 10\%$ for values $b \to 0$. In general, it believes that hadrons slowly turns into BEL (Blacker -- Edger -- Larger) with the increase of $s$ \cite{Cheng-book-1987,Barone-book-2002}. However, some recent experimental results and model approaches, briefly discussed below, indicate this picture is not accurate enough. Thus, the situation is quite complicated as well as far from an unambiguous physical conclusion. The new experimental and phenomenological studies will be crucially important for a deeper understanding of the geometry of the interaction region concerning the hadronic, in particular, $pp$, $\bar{p}p$ collisions.

From the original hypothesis pointed out in
\cite{dremin_1,dremin_2} and later in \cite{broniowski_arriola},
the gray area, also known as hollowness effect, suggests that at
very high energies the inelastic profile function at zero impact
parameter does not reach the unit. This unexpected behavior was
subject of a series of explanations
\cite{broniowski_arriola,alkin_martynov,anisovich_nikonov,dremin_1,troshin_tyurin_1,anisovich,troshin_tyurin_2,albacete_sotoontoso,arriola_broniowski,dremin_2}.
However, none of these approaches took into account the entropy of
the internal constituents of the hadron and associated with the
elastic scattering in the impact parameter space. As shown in
\cite{sdcvaocvm}, the Tsallis entropy (TE) can be connected with
the inelastic overlap function and with the squared critical
energy $s_c$, associated with a phase transition in the
internal structure of (anti)proton which manifests itself via the
total cross section dataset. This phase transition is of topological order, just like the BKT phase transition occurring in the XY--model \cite{sdcvaocvm}. In this kind of phase transition, there is no spontaneous symmetry breaking but only a rearrangement of the internal constituents to some more favorable geometry configuration. This critical value divides the total
cross section experimental data into two samples with different
fractal dimensions, exposing the presence of a multifractal
character in this physical quantity \cite{bc}. It is important to
stress here that the TE emerges exactly in the context of
multifractal structures \cite{tsallis_multifractal}.

On the other hand, the multifractal feature also occurs in the momentum space \cite{antoniou_1,antoniou_2,antoniou_4,antoniou_5,bialas_1,bialas_2}, reinforcing the necessity of using the TE, as shown in \cite{deppman_phys_rev_d93_054001_2016}. Thus, the multifractality of $pp$ and $\bar{p}p$, in different variables, should be noted in the impact parameter space, revealing some novel physical aspect, for example, in the behavior of the inelastic overlap function.

To the complete development of the present approach, one needs the calculation of the internal energy of the hadron. However, this is a very hard, and presently, an unsolved question. Thus, at first glance, one proposes the study of two non-relativistic potentials mimicking the (internal) energy of the hadron in the impact parameter space. The first one is the Coulomb potential being used to represent the total energy of the hadron, which results in the $pp$ and $\bar{p}p$ elastic scattering as billiard ball collisions. Of course, this potential is unable to present results about the internal structure of the hadrons supposed structureless in this case. The approximations performed prevents the extraction of any information near the forward direction ($b < b_{\scriptsize{\mbox{min}}}$). The second approach uses the confinement potential and, contrary to the first one, it allows the presence of an internal structure due to the quarks and gluons. The consequence of the confinement potential, in this case, we claim here, is the arising of the hollowness effect near the forward direction, as shall be seen in this work.

It is important to stress that, in the present thermodynamic approach, we are not interested in presenting a model that produces the best fittings result of the experimental data. The interest here is to study the physical consequences of a phase transition occurring in the $pp$ and $\bar{p}p$ elastic scattering from the TE in the impact parameter space.

The paper is organized as follows. In section \ref{he} one sets
the problem, and a brief review is presented for the experimental situation with absorption in the $pp$ and $\bar{p}p$ collisions. Section \ref{pot} is devoted to general statements, and formalism is introduced, allowing interrelations between thermodynamic quantities, in particular, the internal energy, and the effective potentials in some quantum field theories. In section \ref{sec:4} the Coulomb potential is considered. The possible manifestation of the hollowness effect in the strong interaction, in particular, in $pp$ and $\bar{p}p$ collisions is presented in section \ref{sec:5} with help of the confinement potential. Finally, section \ref{re} contains the discussions and conclusions.

\section{\label{he}Hollowness Effect and Entropy} 

The high energy $pp$ and $\bar{p}p$ elastic scattering can be analyzed using both the transferred momentum $q^2$ or the impact parameter $b$ since these variables can be connected through a Fourier--Bessel transform. Thus, the physical constraints of one space can be rewritten in another one, sometimes revealing details or furnishing insights to solve a problem.

\subsection{Inelastic overlap function}

The impact parameter space is the geometric scenario of the
elastic scattering, possessing clear classical appeal. In this
sense, it is natural to use the black disk picture to describe the
$pp$ and $\bar{p}p$ elastic scattering at high energies. The
elastic scattering amplitude $F(s,q^{2})$ is written using the
impact parameter $b$ as
\begin{eqnarray}
\label{int_eq_1}F(s,q^2)=i4\pi{\int_0}^{\infty}bdbJ_0(qb)\bigl\{1-\exp[i{\chi}(s,b)]\bigr\}=
i4\pi{\int_0}^{\infty}bdbJ_0(qb)\mathrm{\Gamma}(s,b),
\end{eqnarray}

\noindent and here $\sqrt{s}$ is the collision energy in the center-of-mass system, $J_{0}(x)$ is a zeroth order Bessel function, and $\chi(s,b)$ is the eikonal function written as
\begin{eqnarray}
\label{int_eq_2}\chi(s,b)=\mathrm{Re}\chi(s,b)+i\mathrm{Im}\chi(s,b),~~~ \mathrm{Im}\chi(s,b)\geq 0,
\end{eqnarray}

\noindent where the imaginary part corresponds to the opacity
function, identified with the matter distribution inside the
incident particles. It is used the following view of the optical
theorem
$s^{-1}\mathrm{Im}F(s,0)=\sigma_{\scriptsize{\mbox{tot}}}(s)$,
where $\sigma_{\scriptsize{\mbox{tot}}}(s)$ is the total cross
section for the interaction process \cite{sdcvaocvm}.

The profile function
$\Gamma(s,b)=\mathrm{Re}\Gamma(s,b)+i\mathrm{Im}\Gamma(s,b)$ is
the elastic scattering amplitude in the impact parameter space for
$s$ fixed, being able to give an estimate of the particle
interacting radius as well as an glance of its internal structure.
Furthermore, the unitarity condition can be written using the
inelastic overlap function as
\begin{eqnarray}
\label{int_eq_3}0\leq G_{\scriptsize{\mbox{inel}}}(s,b)=2\mathrm{Re}\Gamma(s,b)-|\Gamma(s,b)|^2 \leq 1,
\end{eqnarray}

\noindent that represents the absorption probability of a given $(s,b)$ and, one expects that moving away from the forward collision $(b=0)$, the interaction probability diminishes. The general belief is that, at $b=0$ and for a sufficient high energy, $G_{\scriptsize{\mbox{inel}}}(s,0)\to 1$. This result implies that for $s\to\infty$ the $pp$ and $\bar{p}p$ elastic scattering tends to present the same physical behavior in the forward direction. This behavior is expected to occur due to the leading pomeron exchange \cite{S.D.Campos.Phys.Scr.95.065302.2020,S.D.Campos.arXiv.2003.11493.2020}.

Several theorems were proven from the 1960s, and some of them established the fundamental theoretical basis of the high-energy elastic scattering. For instance, the elastic to total cross section ratio is one of these remarkable results obtained from a well-established basis \cite{froissart_1961,lukaszuk_1967,martin_2009,wu_2011,PRD-91-076006-2015}. The detailed analysis of this cross section ratio in nucleon-nucleon collisions \cite{arXiv-1805.10514} shows that, at high energies, the approximate relation
\begin{eqnarray}
\label{int_eq_4}\sigma_{\scriptsize{\mbox{el}}}(s) / \sigma_{\scriptsize{\mbox{tot}}}(s) \approx 1/4
\end{eqnarray}

\noindent holds for $pp$ up to the highest experimentally reached
energy $\sqrt{s}=57$ TeV within the error bars for this
experimental data. Based on the measured $\sigma_{\scriptsize{\mbox{el}}}(s) / \sigma_{\scriptsize{\mbox{tot}}}(s)$ the quantity $G_{\scriptsize{\mbox{inel}}}(s,0)$ was estimated in unprecedented wide energy range \cite{sdcvaocvm}. These estimations agree reasonably with the model-dependent results obtained some earlier \cite{NPB-166-301-1980,PLB-132-443-1983,PLB-160-167-1985}, but the accuracy of the method from \cite{sdcvaocvm} is usually worse than that for the approach based on the parameterization of the scattering amplitude \cite{NPB-166-301-1980}. The accuracy is $\sim 10\%$ for the method from \cite{sdcvaocvm} and does not allow rigorous and unambiguous conclusions. On the other hand, within the last approach, the evolution of $G_{\scriptsize{\mbox{inel}}}(s,0)$ on $s$ can be studied considering whole experimental energy range available, as commented above. It seems to be an important advantage for the method from \cite{sdcvaocvm}. The intermediate-energy estimations for $G_{\scriptsize{\mbox{inel}}}(s,0)$ in nucleon-nucleon scattering can be approximated by assuming $G_{\scriptsize{\mbox{inel}}}(s,0) \sim 0.9$ constant and some growth is only obtained for $\sqrt{s} > 100$ GeV. As seen, the probability of absorption is close to unity within accuracy $\sim 10\%$ for central collisions at intermediate energies already. The constant behavior of $G_{\scriptsize{\mbox{inel}}}(s,0)$ agrees quite well with the results for the Intersecting Storage Rings (ISR) energy range \cite{NPB-166-301-1980}. The noticeable increase $s$ from the $\sqrt{s}=52.8$ GeV for $pp$ up to the	$\sqrt{s}=546$ GeV for $\bar{p}p$ leads to the corresponding growth of the inelastic overlap function in central collisions from $G_{\scriptsize{\mbox{inel}}}(s,0) \approx 0.9264$ \cite{PLB-132-443-1983} up to the $G_{\scriptsize{\mbox{inel}}}(s,0) \approx 0.9764$ for $\bar{p}p$ interactions \cite{PLB-160-167-1985} indicating the slight growth of the probability of absorption. The increase is confirmed based on a significantly richer experimental material discuted elsewhere \cite{sdcvaocvm}, and it provides a constant $G_{\scriptsize{\mbox{inel}}}(s,0) \sim 1.0$ starting with $\sqrt{s}=546$ GeV up to the highest experimentally reached energy $\sqrt{s}=57$ TeV. In fact results from \cite{sdcvaocvm} confirm the general statement made in Sec. \ref{sec:1} that the (anti)proton turns blacker with the increase of $s$. This statement agrees with the estimations of $G_{\scriptsize{\mbox{inel}}}(s,0)$ for $pp$ collisions at $\sqrt{s}=7$ TeV \cite{PRD-89-051901-2014} indicated by the reaching of the black disk limit and obtained within the modified method from \cite{NPB-166-301-1980}. The black spot appears at the LHC energies $\sqrt{s} \sim 10$ TeV in the central region for both the mode of the (black) partonic disk and the mode of the resonating partonic disk \cite{anisovich}. On the other hand, the mass squared approach with central optical potential shows a shallow minimum at $b=0$ with depth of few percents of the maximum $G^{\scriptsize{\mbox{max}}}_{\scriptsize{\mbox{inel}}}(s,b)=1$, whereas the maximum shifts to $b > 0$ at the LHC energies $\sqrt{s}=7$ and 14 TeV \cite{arriola_broniowski}. This result is corroborated by the observation of a hollowness, i.e. a shallow minimum with depth $\sim 1\%$ of the maximum at $\sqrt{s}=13$ TeV within the dipole Regge model \cite{PRD-98-074012-2018} and the L\'evy imaging \cite{T.Csorgo.R.Pasechnik.A.Ster.Euro.Phys.J.C80.126.2020}. In the last case, the analysis of the high statistic data at $\sqrt{s}=13$ TeV \cite{EPJC-79-861-2019} allows the observation of the proton hollowness effect with beyond 5 s.d. (standard deviation) significance. The analytic extrapolation for $G_{\scriptsize{\mbox{inel}}}(s,0)$ calculated for  an ultra-high energy range in \cite{sdcvaocvm} predicts the onset of deviation from the black disk limit at $\mathcal{O}$(100 TeV), and the continuing decreasing of the inelastic overlap function in central nucleon-nucleon collisions with the growth of $s$ provides $G_{\scriptsize{\mbox{inel}}}(s,0) \to 0$ for PeV energies. In particular, the PeV energy domain is considered as the low boundary for the collision energy for the onset of the noticeable difference between two modes of the partonic disk, black and resonating ones \cite{anisovich}. One can note that the toroid shape of the inelastic interaction region is also evaluated within the assumption that $\mathrm{Im}\,F(s,q^{2}) > 0$ everywhere, with deviation of $G_{\scriptsize{\mbox{inel}}}(0)$ from unit at level of few percentages for multi-TeV energy domain $\sqrt{s} \gtrsim 10$ TeV \cite{EPJC-78-913-2018}. Thus, this brief review confirms the current complex and ambiguous situation in theoretical and experimental studies about the geometry of the hadronic collisions, mostly for $pp$ and $\bar{p}p$.

The unitarity equation (\ref{int_eq_3}) also can be written as
\begin{eqnarray}
\label{se_5}G_{\scriptsize{\mbox{inel}}}(s,b)=\mathrm{Re}\Gamma(s,b)[2-\mathrm{Re}\Gamma(s,b)]-\mathrm{Im}^{2}\Gamma(s,b).
\end{eqnarray}

That result can be rewritten taking into account derivative dispersion relations as well as the crossing property to $\mathrm{Im}\Gamma(s,b)$. Notice that derivative contributions depend on the transferred momentum range. Thus, for $q^2 \to 0$, the derivative contribution occurs in the periphery, while for $q^2 \to \infty$, the contribution is central. Then, the expression (\ref{se_5}) possess two regimes, depending on $b$ is central or peripheral. Considering large values of $b$, $\mathrm{Im}\Gamma(s,b)\approx \mathrm{Re}\Gamma(s,b)$ and, therefore, derivative terms should be taken into account. However, for small values of $b$, $\mathrm{Im}\Gamma(s,b) \ll \mathrm{Re}\Gamma(s,b)$, and derivative terms can be neglected. Considering only small values of $b$, one writes
\begin{eqnarray}
\label{se_6}G_{\scriptsize{\mbox{inel}}}(s,b) \approx \mathrm{Re}\Gamma(s,b)[2-\mathrm{Re}\Gamma(s,b)],
\end{eqnarray}

\noindent and one can identify $1\leq \mathrm{Re}\Gamma(s,b)\leq 2$, where $\mathrm{Re}\Gamma(s,b)=2$ is the completely non-absorptive case and $\mathrm{Re}\Gamma(s,b)=1$ is the full absorptive case. Taking the partial derivative with respect to $b$ of (\ref{se_6}), one obtains ($\partial_{x} \equiv \partial/\partial x$)
\begin{eqnarray}
\label{se_7}\partial_{b} G_{\scriptsize{\mbox{inel}}}(s,b) \approx 2\partial_{b} \mathrm{Re}\Gamma(s,b)[1-\mathrm{Re}\Gamma(s,b)].
\end{eqnarray}

Note that at some critical value $b_c$, the $\left.\partial_{b} G_{\scriptsize{\mbox{inel}}}(s,b)\right|_{b=b_{c}}=0$ can be reached if $\left.\partial_b \mathrm{Re}\Gamma(s,b)\right|_{b=b_{c}} = 0$ and\,/\,or $\mathrm{Re}\Gamma(s,b_{c})=1$. The $and$ connective means that $b_{c}$ is a critical value and the process is completely absorptive at $b_{c}$. On the other hand, the $or$ case is analyzed as follows. If $\left.\partial_b \mathrm{Re}\Gamma(s,b)\right|_{b=b_{c}}=0$ but not $\mathrm{Re}\Gamma(s,b_{c})=1$, then $b_{c}$ is a critical value not representing the full absorptive case, i.e. the inelastic overlap function does not produce the black disk pattern at $b_{c}$. On the contrary, the full absorptive case does not represent a critical point of $\mathrm{Re}\Gamma(s,b)$. This is the non-physical result since the inelastic overlap function is limited. Thus, there are two situations able to furnish a zero in $\partial_b G_{\scriptsize{\mbox{inel}}}(s,b)$ at some impact parameter critical value, $b=b_{c}$. The first situation can be achieved considering that at $b=b_{c}$ the $\left.\partial_b \mathrm{Re}\Gamma(s,b)\right|_{b=b_{c}}=0 \bigcap \mathrm{Re}\Gamma(s,b_c)=1$, hence $b_{c}$ is a critical value and represents the full absorptive case. The second situation can be achieved if $\left.\partial_b \mathrm{Re}\Gamma(s,b)\right|_{b=b_{c}}=0 \bigcap \mathrm{Re}\Gamma(s,b_c)\neq 1$.

Taking into account the allowable range for $\mathrm{Re}\Gamma(s,b)$, then the sign of $\partial_b \mathrm{Re}\Gamma(s,b)$ determines the sign of $\partial_b G_{\scriptsize{\mbox{inel}}}(s,b)$. Considering $\partial_b G_{\scriptsize{\mbox{inel}}}(s,b) > 0$, the only possible physical result is $\partial_b \mathrm{Re}\Gamma(s,b) < 0$ and vice versa. Then, the sign of $\partial_b G_{\scriptsize{\mbox{inel}}}(s,b)$ is controlled only by the sign of $\partial_b \mathrm{Re}\Gamma(s,b)$, and the inelastic overlap function change its sign in agreement with $b$ (fixed $s$). As stated above, $\mathrm{Re}\Gamma(s,b)$ is related to the imaginary part of $F(s,q^2)$ and, changing the sign of $\mathrm{Im}F(s,q^2)$, this also represents a changing in the sign of $\mathrm{Re}\Gamma(s,b)$. As well-known, $\mathrm{Im}F(s,q^2)$ oscillate according to $q^2$ and, therefore, the sign change of $\partial_{q} \mathrm{Im}F(s,q^2)$ occurs as $q^2$ grows.

\subsection{The Tsallis entropy approach}

On the other hand, one can analyze the behavior of
$G_{\scriptsize{\mbox{inel}}}(s,b)$ considering the TE. Notice
that exists several ways to compute the entropy of a thermodynamic
system, being the well-known Boltzmann entropy the most popular.
This entropy is applied, usually, into a system of non-interacting
particles. Hence, this entropy is additive: the entropy of the
whole system is the sum of each subsystems entropy. However, a
system containing interacting subsystems (sometimes strongly
correlated) needs an entropy calculation that takes this feature
into account.

The TE can naturally be applied into correlated systems since it is non-additive. Moreover, R\'enyi entropy \cite{renyi_entropy}, Shannon entropy \cite{shannon_entropy}, Abe entropy \cite{abe_entropy} and Boltzmann entropy, for instance, can be reduced to the TE \cite{beck_0902.1235v2, tsallis_book}. Furthermore, the TE possesses two (among others) interesting mathematical properties: it is concave for all $w>0$, a crucial characteristic for an entropy function. Besides, it also obeys the Lesche stability condition, i.e. it is stable under small perturbations of probabilities. Considering these properties, the TE is able to furnish a description of the physical system under study.

Bearing in mind the above considerations the TE entropic index $w$ can be replaced by the ratio \cite{sdcvaocvm}
\begin{eqnarray}
\label{se_1} w=s/s_{c},
\end{eqnarray}

\noindent where $s_{c}$ is the squared critical energy associated with the BKT-like phase transition \cite{sdcvaocvm}, whose consequence is the fractal structure of the total cross-section \cite{bc}. In this sense, $w$ plays the physical role of a transition parameter. When $s > s_{c}$, the fractal dimension is positive and negative when $s < s_{c}$, i.e. the TE possess two behaviors depending on $w > 1$ or $w < 1$. On the other hand, the negative fractal dimension can be viewed as a measure of the hadron emptiness (the slowdown part of the total cross-section data set); the positive fractal dimension can be associated with the usual measure of the total cross section (the growing part of the total cross-section data) \cite{bc}.

In \cite{sdcvaocvm}, the TE is identified with the scattering at
fixed $s$ by means of the inelastic overlap function
$G_{\scriptsize{\mbox{inel}}}(s,b)$, due to non-elastic
$s$-channel intermediate states as ($k \equiv s/s_{c}-1$)
\begin{eqnarray}
\label{se_2} S_T\equiv S_T(s,b)=k^{-1}[1-G_{\scriptsize{\mbox{inel}}}(s,b)],
\end{eqnarray}

\noindent where the probability of an event in the impact
parameter space $\displaystyle
P(s,b)=\int_{0}^{b}\bigl[p\,(b')\bigr]^{s/s_{c}}db'$ is replaced
by $G_{\scriptsize{\mbox{inel}}}(s,b)$ within the hypothesis for
using of the TE in the $b$-space and, for the sake of simplicity,
one assumes $m=1$ and $n=1$ \cite{sdcvaocvm}. It is interesting to
note that unitarity demands $0\leq
G_{\scriptsize{\mbox{inel}}}(s,b)=1-k S_T \leq 1$ implying the
replacement $k S_T \rightarrow\tilde{S}_{T}$, where
$\tilde{S}_{T} \equiv kS_{T}$ is the normalized entropy.

As well-known, the inelastic overlap function takes into account all intermediate inelastic channel contributions. Thus, the entropy (\ref{se_2}) can be associated with the inelastic scattering contributions. In the above result, if $G_{\scriptsize{\mbox{inel}}}(s,b) \to 1$ (the black disk limit) then $S_T \to 0$. The physical meaning of this result is simple: at the black disk limit, the system (the motion of the internal constituents) is in its lowest (or highest) possible value, as stated by Quantum Mechanics. Therefore, the physical state of the system is well defined.

The inelastic overlap function is interpreted as the probability
of an inelastic scattering in a given $(s,b)$. Thus,
$G_{\scriptsize{\mbox{inel}}}(s,b)=1$ implies that at head-on
collision $b=0$ (or at some $b\neq 0$ as professed by the
hollowness effect), the probability achieves its maximum as well
as the entropy tends to its minimum. The general belief is that
when $s \to \infty$ the black disk limit is achieved at $b=0$.
However, this is not necessarily true since there is a sign change
in the TE in accordance with $s/s_{c}$. To see this, observe
that partial derivative of $S_{T}$ with respect to $b$ is given by
\begin{eqnarray}
\label{se_3}\partial_{b} S_{T}(s,b)=-k^{-1}\partial_{b} G_{\scriptsize{\mbox{inel}}}(s,b).
\end{eqnarray}

Assuming $s>s_c$ (high energies regime), the sign of $\partial_b
S_T$ is determined by the sign of $\partial_b
G_{\scriptsize{\mbox{inel}}}(s,b)$. In this regime, the fractal
dimension of the total cross section is positive representing the
matter density increase inside the hadron \cite{bc}. Therefore, in
accordance with the above analysis, $b_c$ determines the region
inside the hadron where the entropy increases ($b>b_c$) or
decreases ($b<b_c$). On the other hand, considering $s<s_c$ (low
energies regime), the existence of $b_c$ implies in an increasing
($b<b_c$) or decreasing ($b>b_c$) entropy. The fractal dimension
of the total cross section is negative, representing the emptiness
or the absence of a well-defined internal structure inside the
hadron \cite{bc}. The same result can be obtained replacing
(\ref{se_7}) into (\ref{se_3}), showing a matter distribution in
accordance with the existence of $b_c$ and determining the entropy
behavior.

\section{\label{pot}Internal Energy and Effective Potential} 

Assuming statistical equilibrium between a heat reservoir with the
temperature $\Theta$ and a hadron the later can be considered as
the canonical ensemble of its constituents at temperature $T$ and
consequently for nonextensive statistics \cite{PA-261-534-1998}
\begin{eqnarray}
\label{eq:3.1} \partial_{\mathcal{U}}\tilde{S}_{T} =
\tilde{T}^{-1},
\end{eqnarray}

\noindent where definition of the canonical ensemble
\cite{Levich-book-1971,JIKapusta-GCharles-book-2006} or,
equivalently, closed system \cite{kaufman_book_2001}  is taken
into account, $\mathcal{U}$ is the hadron internal energy,
$\tilde{T} \equiv T/k$ is the normalized temperature of the
constituent ensemble under consideration, $\Theta=T$
\cite{Levich-book-1971} and $T=T(s,b)$ due to corresponding
dependences of ${S}_{T}$.

In consonance with the preceding section, $S_T$ is replaced by its
normalized form since $G_{\scriptsize{\mbox{inel}}}(s,b)$ obey the
unitarity condition. Therefore, the entropy of the above system of
constituents also can be written using the thermodynamics. The
approach of the canonical ensemble or, more generally, grand
canonical one at negligibly small chemical potential, allows the
suggestion a constant Helmholtz free energy ($\mathcal{F}$).
Therefore the (\ref{eq:3.1}) can be rewritten in the standard
integral form $\mathcal{U}=\mathcal{F}+\tilde{T}\tilde{S}_{T}$ in
which the constant of integration is assigned as $\mathcal{F}$.
The quantity $\mathcal{U}$, as well-known, cannot be deduced from
the first principles of thermodynamics. For quantum systems
studied here $\mathcal{U}$ can be reduced to the potential energy
$E_{p}$ \cite{Shuryak-book-2004-p374} which is a function of the
\textit{effective} potential $V$, thus $\mathcal{U}\approx
E_{p}(V)$. Of course, this rough approximation excludes the
kinetic term and the action due to external forces. Taking into
account $V(s,r)=E_{p}(s,r)-E_{p}(s,\infty)$ \cite{PRD-70-054507-2004} one can derive
$V+E_{p}(s,\infty) \approx \mathcal{F}+\tilde{T}\tilde{S}_{T}$,
where $r$ is the distance between constituents. For an Abelian
quantum field theory (QFT), like QED, $E_{p}(s,\infty)=0$. The
hadron as a statistical system will spontaneously undergo a
process if it lowers the systems Helmholtz free energy
\cite{kaufman_book_2001}. Thus the ordinary hadron should be
characterized by the lowest value of $\mathcal{F}$ which can be
assigned as the zero (ground) level. On the other hand, in some
non-Abelian QFT, like QCD in the pure gauge limit, $E_{p}(s,r \to \infty) \to \infty$ and
the Helmholtz free energy tends to the infinitely large value at
$r \to \infty$ within the approach of a static constituents at temperature smaller than $T$ of a phase transition\footnote{Within QCD in the pure gauge limit this situation corresponds to the contribution of static (anti)quarks with a infinite mass to the heat reservoir \cite{AIPCP-602-323-2001}}. Such cold system can be considered as a stationary confinement state, i.e. as (quasi)hadron in the strong interaction field. Moreover
similar growth can be suggested for $E_{p}$ and $\mathcal{F}$ at
increasing of $r$ for any distances at negligibly small $T$ based on the available results for finite values of $r$ obtained with help of the lattice QCD calculations
\cite{AIPCP-602-323-2001,PTPSuppl-153-287-2004} as well as the phenomenological studies, in particular,
within $T$-matrix formalism \cite{NPA-941-179-2015}. Therefore assuming a mutual reduction of the
terms $E_{p}(s,\infty)$ and $\mathcal{F}$, at least, qualitatively
as well as an appropriate replacement $r \to b$ the following
general relation can be deduced for a stationary state
\begin{eqnarray}
\label{eq:3.1.add}V(s,b) \approx \tilde{S}_{T}(s,b)\tilde{T}(s,b),
\end{eqnarray}

It should be stressed that the Bohm's quantum potential can be
used to mimic the internal energy of a quantum system, giving
insight into its role in stationary states
\cite{G.Dennis.M.A.de.Gosson.B.J.Hiley.Phys.Lett.A378.2363.2014}.
Then, in the Bohm's point of view, the particle is not a
point-like object, contrariwise, it possesses an internal
structure with some topological geometry. As shall be seen, this
extended structure is necessary to explain the hollowness effect.

The temperature must be normalized to obey the unitarity condition. On the other hand, the relevant information here is the sign of the temperature, depending on $s$, since this approach (the use of the effective potential) does not allows the precise knowledge of the critical temperature. Hence, normalizing the temperature one still maintain the relevant information about its sign only by using the procedure
\begin{eqnarray}
\tilde{T}(s,b)=\left\{
\begin{array}{lcl}
\displaystyle +1, & \mbox{if} & s<s_c\\
\displaystyle -1, & \mbox{if} & s>s_c\\
\end{array}\right.\label{eq:3a.1c}
\end{eqnarray}

\noindent Mathematically speaking, the only requirement to obtain a negative temperature is that the entropy should not be restricted to monotonically increasing of $\mathcal{U}$ \cite{ramsey_phys_rev_103_10_1956}. Its physical meaning is also well-defined: the occupation distribution is inverted, where high-energies states are populated more than low-energies states. The occupation probability of a quantum state increases exactly with the energy of the state. Keeping the information about the phase transition, the qualitative behavior of the inelastic overlap function is studied here. Based on the (\ref{eq:3.1.add}) the following chain of the equations can be obtained within the potential approach: $\partial_b \tilde{S}_{T} = \tilde{T}^{-1}(s,b) \partial_b V(s,b) - V(s,b) \tilde{T}^{-2}(s,b) \partial_b \tilde{T}(s,b)= \tilde{T}^{-1}(s,b) \partial_b V(s,b)$ taking into account (\ref{eq:3a.1c}). Then, it is deduced the particular relation $\partial_b G_{\scriptsize{\mbox{inel}}}(s,b)=-\tilde{T}^{-1}(s,b)\partial_b V(s,b)$, in which one can use $|\tilde{T}(s,b)|$ without lost of generality. It allows the use of a simple {\it ansatz}
\begin{eqnarray}
\label{eq:3.2}G_{\scriptsize{\mbox{inel}}}(s,b)=1-\tilde{V}(s,b),
\end{eqnarray}

\noindent to solve the last differential equation within the
potential approach, where $\tilde{V}(s,b)$ can be obtained with
the help of some procedure from the potential $V(s,b)$ in order to
preserve the validity of the unitarity condition (\ref{int_eq_3}).
Taking into account this condition, then $0 \leq \tilde{V}(s,b)
\leq 1$ and, consequently, the normalization can be suggested as
such procedure with the specific details depending on the view and
behavior of the $V(s,b)$ in the kinematic region under study. It
should be noted that there are some restricted ranges for
the impact parameter ($b_{\scriptsize{\mbox{min}}} \leq b \leq
b_{\scriptsize{\mbox{max}}}$) and for the collision energy
($s_{\scriptsize{\mbox{min}}} \leq s\leq
s_{\scriptsize{\mbox{max}}}$), since in hadronic interactions
these parameters are characterized by finite values for the
boundaries $b_{\scriptsize{\mbox{min}/\mbox{max}}}$,
$s_{\scriptsize{\mbox{min}/\mbox{max}}}$ due to, in general,
finite space scales ("sizes'') of incoming particles and finite
collision energy for any physical process. The reliable values of the boundaries $b_{\scriptsize{\mbox{min}/\mbox{max}}}$,
$s_{\scriptsize{\mbox{min}/\mbox{max}}}$ are defined within concrete approach used and /or kinematic features of the reaction under consideration. The following general statement can be
obtained from (\ref{eq:3.2}): the black disk regime
$G_{\scriptsize{\mbox{inel}}}(s,b) \to 1$ is reached only if
$\tilde{V}(s,b) \to 0$ in some kinematic domain and / or separate
points of the $(s,b)$ plane. Thus, within the potential approach,
the above {\it ansatz} produce the result
\begin{eqnarray}
\label{eq:3.4}\tilde{S}_{T}(s,b)=\tilde{V}(s,b)
\end{eqnarray}

\noindent replacing (\ref{eq:3.2}) into the relation (\ref{se_2}).
Consequently, the $b$-dependence of the TE is the same as for
effective potential $\tilde{V}(s,b)$. In general, the
$s$-dependence of the $S_{T}$ for certain types of the potential
$V(s,b)$ can be deduced with the help of equation (\ref{eq:3.4})
and $k$, in agreement with the definition of normalized TE and the
appropriate choice of $s_{c}$. In addition, one can note that the
equation (\ref{eq:3.4}) is in accordance with (\ref{eq:3.1})
taking into account the replacement $\mathcal{U} \to V$ and
normalization (\ref{eq:3a.1c}) made above.

Depending on the potential used, this assumption allows or not a view on the internal structure of the particles. One considers here two potentials in the impact parameter space as attempts to explain the behavior of the inelastic overlap function. The first one is the well-known Coulomb potential, which allows a naive view of the inelastic overlap function from the {\it outside} of the hadron. This potential is used for structureless particles. The second one is the confinement potential that represents the point of view of the constituents of the hadron \cite{sdcvaocvm}. One supposes this thermodynamic system is described by the canonical ensemble, where particle exchange is forbidden. Thus, the proton, as well as the antiproton, is a composite particle, turning relevant know how the collision energy is shared among the quarks and gluons. That question is quite similar to the multiplicity scenario and will be discussed further.

\section{Coulomb Potential}\label{sec:4}

At this first moment, one assumes the Coulomb potential as being able to describe the hadron energy treating it as a point-like particle. Despite this naive approach, it can furnish at least a classical picture of the inelastic overlap function. Assuming $r=|\mathbf{r}_i-\mathbf{r}_j|$ as the distance between hadrons placed at $\mathbf{r}_i$ and $\mathbf{r}_j$ and with masses $m_{h}^2\ll s$, then using the impact parameter $b$ one can approximate the Coulomb potential $V_{\scriptsize{\mbox{C}}}(r)=-a/r$ at fixed $s$ by
\begin{eqnarray} \label{eq:3a.1}
V_{\scriptsize{\mbox{C}}}(r) \approx V_{\scriptsize{\mbox{C}}}(s,b)=-ab^{-3}\bigl(b^{2}-2/s\bigr).
\end{eqnarray}

\begin{figure}[ht]
\centering
\includegraphics[width=14.5cm,height=9.5cm]{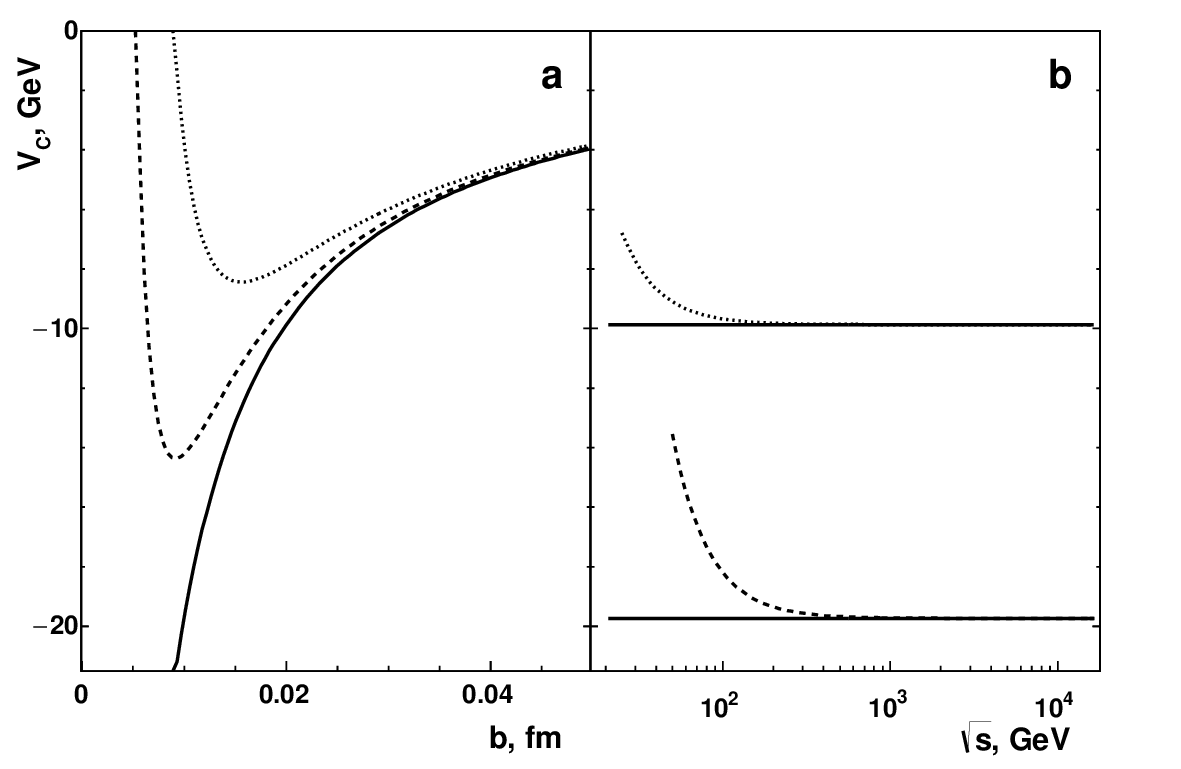}
\caption{\label{fig:1}Dependence of the exact view of Coulomb potential (solid lines) and approximate relation (\ref{eq:3a.1}) on the (a) impact parameter and (b) collision energy: a -- the dotted line corresponds to the approximation (\ref{eq:3a.1}) at $\sqrt{s}=31.0$ GeV and the dashed line -- to the (\ref{eq:3a.1}) at $\sqrt{s}=52.8$ GeV; b -- the dotted curve is the approximation (\ref{eq:3a.1}) at $b=0.02$ fm and the dashed line -- the (\ref{eq:3a.1}) with $b=0.01$ fm. The values $a=1$, $\theta=0$ are used in approximate view of the $V_{\scriptsize{\mbox{C}}}(r)$, for the sake of simplicity.}
\end{figure}

\noindent The above representation can be obtained by noting that $b=r\cos \theta$ is the projection of $r=|\vec{r}|$ onto $b=|\vec{b}|$, where $\theta$ is the angle between $\vec{b}$ and $\vec{r}$, where $r=|\vec{r}|$ is the distance between the hadrons. Moreover, $q^2=2k^2(1-\cos\theta)$ and $s/2-2m^2=2k^2$, where $m$ is hadron mass and $k=|\vec{k}|$ is the norm of the three-momentum \cite{Barone-book-2002}. The parameter $a > 0$ is dimensionless corresponding to the electrostatic interaction of pair. Note the result (\ref{eq:3a.1}) can be written as $V_{\scriptsize{\mbox{C}}}(s,b)\approx-ab^{-1}$ for a sufficient high fixed-$s$. Fig. \ref{fig:1} shows the $b$- (a)
and the $s$-dependence (b) for exact view of Coulomb potential and
its approximation in the impact parameter space (\ref{eq:3a.1}).
In the latter case, the curves are shown for fixed $\sqrt{s}=31.0$
and 52.8 GeV (Fig. \ref{fig:1}a) and for fixed $b=0.01$ and 0.02
fm (Fig. \ref{fig:1}b).

The approximation performed above is better for peripheral collision
than for central, since $V_{\scriptsize{\mbox{C}}}(s,b)\rightarrow
V_{\scriptsize{\mbox{C}}}(r)$ to $b\rightarrow \infty$ and
$r\rightarrow\infty$, as shown in Fig. \ref{fig:1}a\footnote{We
are not interested here in the description of the tail of the
inelastic overlap function. Therefore, we do not take into account
derivative terms.}. The minimum of
$V_{\scriptsize{\mbox{C}}}(s,b)$ is settle down at
$b_{\scriptsize{\mbox{min}}}=\sqrt{6/s}$,
$V_{\scriptsize{\mbox{C}}}(s,b_{\scriptsize{\mbox{min}}})=-a\sqrt{2s/27}$,
which roughly means that considering $b\gtrsim
b_{\scriptsize{\mbox{min}}}$, $V_{\scriptsize{\mbox{C}}}(s,b)
\simeq V_{\scriptsize{\mbox{C}}}(r)$, as seen in Fig.
\ref{fig:1}a. Thus, the approximation done in (\ref{eq:3a.1}) is
used to $b > b_{\scriptsize{\mbox{min}}}$ and no information
considering $b \to 0$ can be obtained, i.e. any information
obtained from $V_{\scriptsize{\mbox{C}}}(s,b <
b_{\scriptsize{\mbox{min}}})$ may not correctly describe the
elastic scattering from the impact parameter point of view. At a
given $b$, the approximate function
$V_{\scriptsize{\mbox{C}}}(s,b)$ will reasonably agree with the
curve for the exact Coulomb potential at $s >
s_{\scriptsize{\mbox{min}}}$, being improved as que collision
energy grows, where
$\sqrt{s_{\scriptsize{\mbox{min}}}}=\sqrt{6}/b$ for fixed $b$.
This statement is confirmed in Fig. \ref{fig:1}b: the range of $\sqrt{s}$ where the accordance between the curves coincide for $V_{\scriptsize{\mbox{C}}}(r)$ and
$V_{\scriptsize{\mbox{C}}}(s,b)$ diminishes on the smaller collision energies and with the growth of $b$.

The Coulomb potential is of long-range and $\forall\, r: V_{\scriptsize{\mbox{C}}}(r) < 0$, consequently, $V_{\scriptsize{\mbox{C}}}(s,b) < 0$ for any $b$ and $s$. Considering the potential with constant sign, the following normalization is used
\begin{eqnarray}
\label{eq:3.3}\tilde{V}(s,b)=\bigl[V^{\scriptsize{\mbox{max}}}(s,b)/V(s,b)\bigr]^{\gamma},
\end{eqnarray}

\noindent where $V^{\scriptsize{\mbox{max}}}(s,b)$ represents its
maximum, $\gamma=\pm 1$ with up (down) sign for negative
(positive) values of $V(s,b)$, within the whole range of the
kinematic parameter values considered. Thus, using the {\it
ansatz} (\ref{eq:3.2}), one writes
\begin{eqnarray}
\displaystyle
\label{eq:3a.2}G_{\scriptsize{\mbox{inel}}}^{\,\scriptsize{\mbox{C}}}(s,b)=1-\tilde{V}_{\scriptsize{\mbox{C}}}(s,b)=1-
\frac{b}{b_{\scriptsize{\mbox{max}}}}\frac{1-2/sb_{\scriptsize{\mbox{max}}}^{2}}{1-2/sb^{2}},
\end{eqnarray}

\noindent where $\tilde{V}_{\scriptsize{\mbox{C}}}(s,b)$ is the
effective (normalized) Coulomb potential defined by (\ref{eq:3.3}) and
taking $\gamma=1$ and
$V_{\scriptsize{\mbox{C}}}^{\scriptsize{\mbox{max}}}(s,b) \equiv
V_{\scriptsize{\mbox{C}}}(s,b_{\scriptsize{\mbox{max}}})$, as a
result of the negative values and smooth behavior of the
$V_{\scriptsize{\mbox{C}}}(s,b)$ shown in Fig. \ref{fig:1}.

The impact parameter $b_{\scriptsize{\mbox{max}}}$ is the appropriate upper boundary value for $b$, and here we use $b_{\scriptsize{\mbox{min}}} \ll b_{\scriptsize{\mbox{max}}}$ for the calculation of $b_{\scriptsize{\mbox{max}}}$. The detailed study of Fig. \ref{fig:1} assumes that $G_{\scriptsize{\mbox{inel}}}^{\,\scriptsize{\mbox{C}}}(s,b)$, defined by (\ref{eq:3a.2}), can only describe the region $b \geq b_{\scriptsize{\mbox{min}}}$, for fixed $s$ and within the range $s \geq s_{\scriptsize{\mbox{min}}}$, for fixed $b$. The approximate relation
\begin{eqnarray}
\label{eq:3a.3} \tilde{V}_{\scriptsize{\mbox{C}}}(b) \approx b /
b_{\scriptsize{\mbox{max}}},
\end{eqnarray}

\noindent is applicable for $b > b_{\scriptsize{\mbox{min}}}$ in the kinematic domain of validity of the condition $sb^{2} \gg 2$. The approximately energy-independent behavior of $G_{\scriptsize{\mbox{inel}}}^{\,\scriptsize{\mbox{C}}}(s,b)$ is expected in almost whole allowed range $b \in [b_{\scriptsize{\mbox{min}}},b_{\scriptsize{\mbox{max}}}]$, with  the exception of a narrow region, close to the lower boundary. The (very) weak dependence on $G_{\scriptsize{\mbox{inel}}}^{\,\scriptsize{\mbox{C}}}(s,b)$ over $\sqrt{s}$ may be caused by the approximate relation (\ref{eq:3a.3}) as well as due to the range considered. Such behavior of $G_{\scriptsize{\mbox{inel}}}^{\,\scriptsize{\mbox{C}}}(s,b)$, can be expanded for larger $b$ with the increase of the boundary value $b_{\scriptsize{\mbox{max}}}$.

\begin{figure}[ht]
\centering
\includegraphics[width=9.0cm,height=9.0cm]{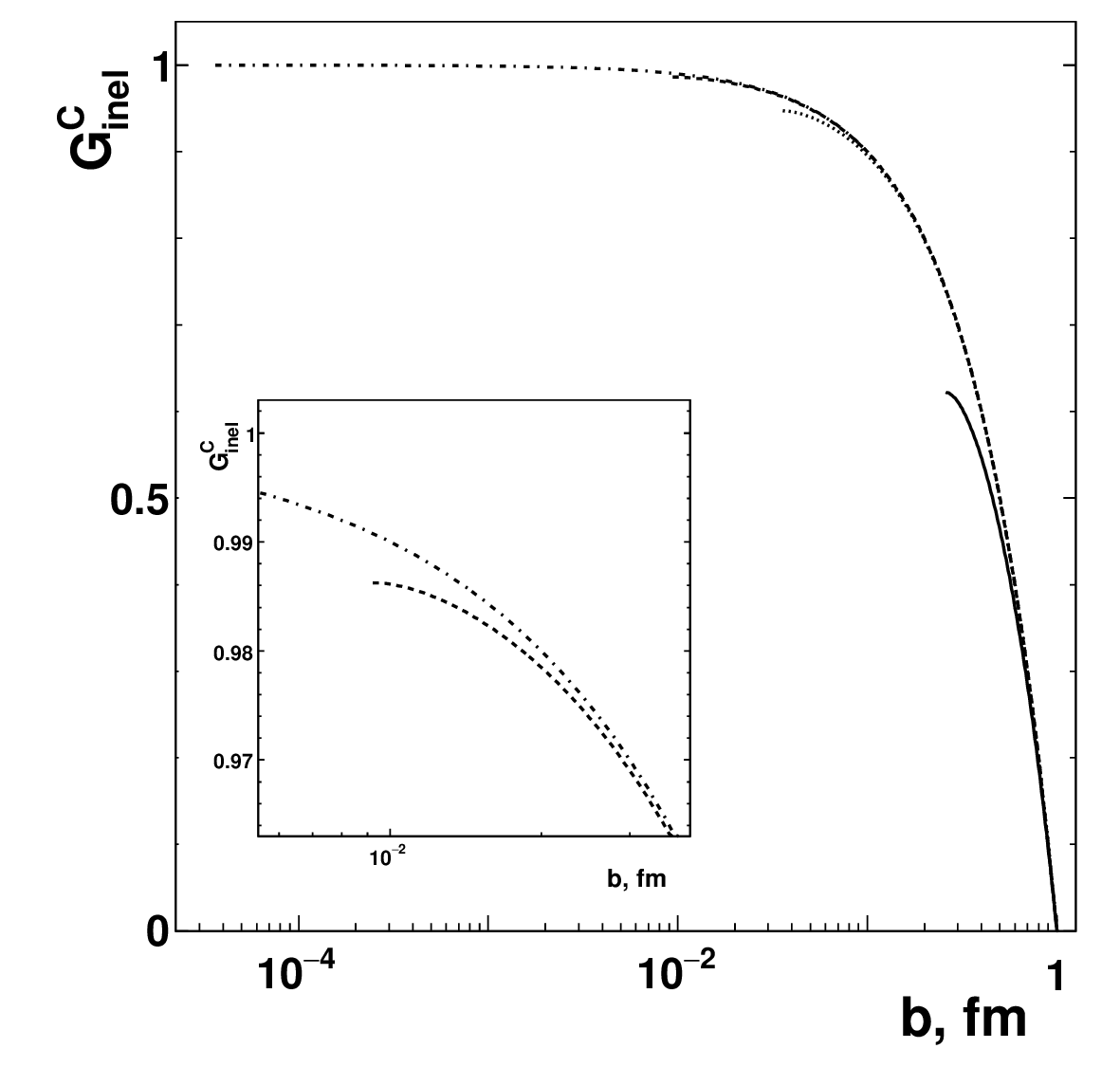}
\caption{\label{fig:2}The behavior of
$G_{\scriptsize{\mbox{inel}}}^{\,\scriptsize{\mbox{C}}}(s,b)$ using the Coulomb potential, where the full line considers
$\sqrt{s_{\scriptsize{\mbox{l.b.}}}} \approx 1.88$ GeV, dotted one $\sqrt{s}=13.8$ GeV, dashed curve $\sqrt{s}=52.8$ GeV and dot-dashed line $\sqrt{s}=14$ TeV. The curves are obtained considering the approximation
(\ref{eq:3a.1}) and $b_{\scriptsize{\mbox{min}}} \leq b \leq b_{\scriptsize{\mbox{max}}}$ with
$b_{\scriptsize{\mbox{max}}}=1.0$ fm. Inner panel: the region of the visible difference between two curves at $\sqrt{s}=52.8$ GeV and $\sqrt{s}=14$ TeV.}
\end{figure}

The Fig. \ref{fig:2} shows the behavior of the inelastic overlap function in accordance with $b_{\scriptsize{\mbox{max}}}$ and for several collision energies from the low-boundary. This boundary is the minimum allowable energy for nucleon-nucleon scattering $\sqrt{s_{\scriptsize{\mbox{l.b.}}}}=2m_{p}$ up to the nominal energy for $pp$ mode at the LHC, where $m_{p}$ is the proton mass \cite{PDG-PRD-98-030001-2018}. The choice $b_{\scriptsize{\mbox{max}}}=1$ fm is a result of the typical linear scale of hadron physics.

As been seen above,
$G_{\scriptsize{\mbox{inel}}}^{\,\scriptsize{\mbox{C}}}(s,b)$ is
characterized by the approximately flat behavior at small $b
\lesssim 0.1$ fm, with consequent fast decreasing as $b$ grows for
the energy range $\sqrt{s} \gtrsim 14$ GeV. Such behavior may be
associated with the absence of an internal structure which is, of
course, a result of the naive potential adopted. The weak changing
region of
$G_{\scriptsize{\mbox{inel}}}^{\,\scriptsize{\mbox{C}}}(s,b)$
narrows with the decreasing of the collision energy.

The inner panel is confirmed and (Fig. \ref{fig:2}) shown the region of the visible difference between two curves $G_{\scriptsize{\mbox{inel}}}^{\,\scriptsize{\mbox{C}}}(s,b)$ at $\sqrt{s}=52.8$ GeV (dashed line) and $\sqrt{s}=14$ TeV (dot-dashed line). These features of the behavior of $G_{\scriptsize{\mbox{inel}}}^{\,\scriptsize{\mbox{C}}}(s,b)$ are in full accordance with a detailed analysis of the relation (\ref{eq:3a.3}).

It is interesting to note that, in the approach of point-like
hadrons, as the collision energy increases,
$G_{\scriptsize{\mbox{inel}}}^{\,\scriptsize{\mbox{C}}}(s,b)$
shown in Fig. \ref{fig:2}, extends to very small values of $b$.
Furthermore, the behavior  of
$G_{\scriptsize{\mbox{inel}}}^{\,\scriptsize{\mbox{C}}}(s,b)$
corresponds to the black disk approach considering
$b_{\scriptsize{\mbox{min}}} < b \lesssim 0.10$ fm, and there is
no signatures for hollowness effect for any $\sqrt{s}$. At
$\sqrt{s} \lesssim 14$ GeV, the inelastic overlap function
decreases with the increase of $b$, in almost the entire allowed
impact parameter range. The value of
$G_{\scriptsize{\mbox{inel}}}^{\,\scriptsize{\mbox{C}}}(s,b)$ is
significantly smaller than 1.0 and the black disk approach is not
valid in the energy range $\sqrt{s} \lesssim 14$ GeV.

The $b$-dependence of the TE on the Coulomb potential can be
immediately derived from the Fig. \ref{fig:2} and relation
(\ref{eq:3a.2}). At qualitative level, the normalized TE, adopting
the Coulomb potential
$\tilde{S}_{T}^{\,\scriptsize{\mbox{C}}}(s,b)$, is characterized
by very small values of $b_{\scriptsize{\mbox{min}}} < b \lesssim
0.10$ fm with fast growth. And
$\tilde{S}_{T}^{\,\scriptsize{\mbox{C}}}(s,b) \to 1$ at large
enough impact parameter values $b \sim
b_{\scriptsize{\mbox{max}}}$, i.e. for peripheral collisions.

\begin{figure}[ht]
\centering
\includegraphics[width=9.0cm,height=9.0cm]{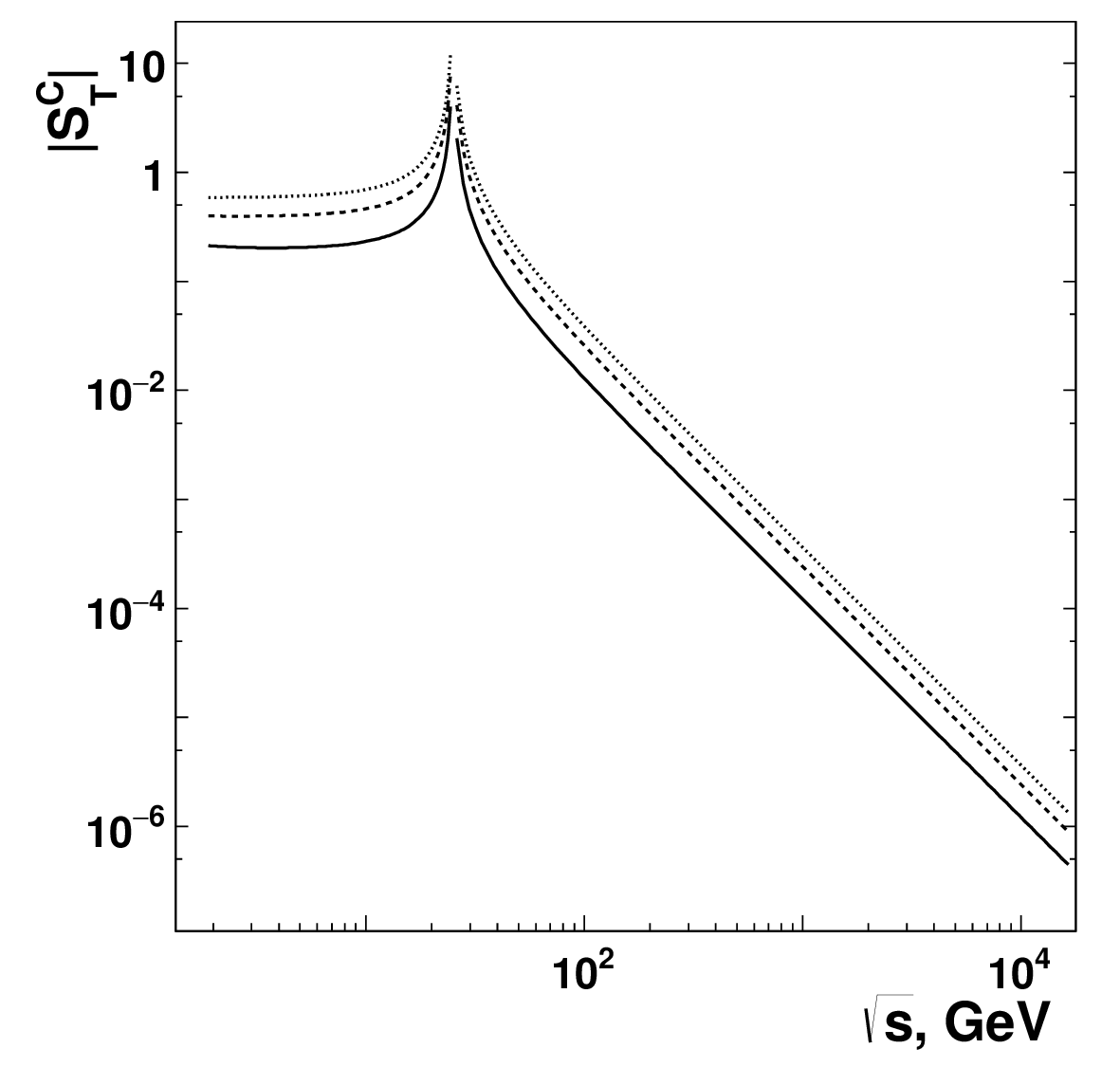}
\caption{\label{fig:3}Absolute values of the TE deduced for the Coulomb potential in the impact parameter space, considering $\sqrt{s_{c}}=25.0$ GeV, $b_{\scriptsize{\mbox{min}}}=\sqrt{6/s_{\scriptsize{\mbox{l.b.}}}}$ and $b_{\scriptsize{\mbox{max}}}=10b_{\scriptsize{\mbox{min}}}$. The solid lines correspond to $b=0.5$ fm, dashed one -- to the $b=1.0$ fm and dotted lines are for $b=1.5$ fm. Details are described in the text.}
\end{figure}

In accordance with the general view shown above, the
$s$-dependence of the TE for the Coulomb potential is deduced by
substituting Eq. (\ref{eq:3a.2}) into the relation (\ref{se_2}).
Considering $\sqrt{s_{c}}=25$ GeV and $b
<b_{\scriptsize{\mbox{max}}}=\varepsilon_{1}b_{\scriptsize{\mbox{min}}}$,
where the $\varepsilon_{1}=10$ and $b_{\scriptsize{\mbox{min}}}$
is defined by $s_{\scriptsize{\mbox{l.b.}}}$, in order to (i) the
condition $b_{\scriptsize{\mbox{min}}} \ll
b_{\scriptsize{\mbox{max}}}$ be correct and (ii) the whole
available energy range for calculation for certain $b$.

The detailed analysis of (\ref{se_2}) reveals that $k$ defines the
sign of the
$\left.S_{T}^{\,\scriptsize{\mbox{C}}}(s,b)\right|_{b=b_{\scriptsize{\mbox{fixed}}}}$,
and this quantity shows a sharp behavior for $s\rightarrow s_{c}$.
Furthermore, the absolute values of the TE for $s < s_{c}$
($|S_{T}^{\,\scriptsize{\mbox{C}}}|=-S_{T}^{\,\scriptsize{\mbox{C}}}$)
are larger by orders of magnitude than that for $s > s_{c}$
($|S_{T}^{\,\scriptsize{\mbox{C}}}|=S_{T}^{\,\scriptsize{\mbox{C}}}$).
Therefore, the $|S_{T}^{\,\scriptsize{\mbox{C}}}|$ seems the more
adequate function for the study of the $s$-dependence of the TE
using the approximation (\ref{eq:3a.1}) for the Coulomb potential
in $b$-space.

The energy dependence on the $|S_{T}^{\,\scriptsize{\mbox{C}}}|$
is shown in Fig. \ref{fig:3} for several values of $b$. As
expected, the $|S_{T}^{\,\scriptsize{\mbox{C}}}|(s)$ is
characterized by a sharp behavior close to the critical energy
$s_{c}$, with a subsequent smooth decrease. The
$|S_{T}^{\,\scriptsize{\mbox{C}}}|$ assume finite values for $s
\to s_{\scriptsize{\mbox{l.b.}}}$ in the energy domain $s <s_{c}$.
On the other hand, the absolute value of the TE (\ref{se_2}), for
the Coulomb potential, decreases significantly with the collision
energy growth for $s > s_{c}$ (Fig. \ref{fig:3}).

The energy dependence on $|S_{T}^{\,\scriptsize{\mbox{C}}}|$ is
mostly defined by the $k$-factor. The influence of the
$\tilde{V}_{\scriptsize{\mbox{C}}}(s,b)$ is weak and it only
manifests itself at low and intermediate energies: at the low
boundary $s_{\scriptsize{\mbox{l.b.}}}$. The relative difference
between the exact $\tilde{V}_{\scriptsize{\mbox{C}}}(s,b)$ and the
$s$-independent approximation (\ref{eq:3a.3}) is about 9\% for
$b=0.5$ fm and $\simeq 0.7$\% for $b=1.5$ fm. Moreover, this
difference decreases fast as the energy growths and it is
negligible ($< 0.5$\%) for $\sqrt{s} \lesssim 8$ GeV for any
considered $b$.

The Coulomb scattering treats the hadrons as billiard balls and
does not take into account the influence of the internal
arrangement of quarks and gluons for the complete description of
the total cross section (or any physical observable). Therefore,
any analysis of the elastic scattering should take into account
quarks and gluons, which may avoid the occurrence of
$b_{\scriptsize{\mbox{min}}}$, presenting a physical explanation
of what occurs for $b < b_{\scriptsize{\mbox{min}}}$.

\section{Confinement Potential}\label{sec:5}

For definition the hadron is considered here as the cold system of
quarks and gluons, i.e. as the quark-gluon matter at $T \ll
T_{\chi, c}$, where $T_{\chi}$ is the temperature for the chiral
symmetry restoration, $T_{c}$ is the temperature for the
confinement transition and $T_{\chi, c} \approx 0.15-0.16$ GeV. At
such negligibly small temperatures it is customary to obtain the
confinement potential ($V_{\scriptsize{\mbox{c}}}$) by adding a
linear term to the Coulomb--like potential. Thus, the
Coulomb--like part responds by the weak interaction of the
antiquark-quark ($\bar{q}q$) pair at short distances while the
linear term describes the strong interaction at large distances,
i.e. the nature of confinement. As indicated above one supposes
the system can be described by the canonical ensemble. In the
lowest order, the confinement potential can be written as
\cite{PRD-17-3090-1978}
\begin{eqnarray}
\label{eq:3b.1} V_{\scriptsize{\mbox{c}}}(\mu,r) = -4\alpha_s(\mu)/3r+\kappa r,
\end{eqnarray}

\noindent where $r$ is now the spatial separation of the pair,
strictly speaking, the infinitely heavy (static) quarks and
antiquarks inside the hadron. The running coupling constant
$\alpha_{s}(\mu)$ is responsible by the strong interaction for a
specific energy scale $\mu$ \cite{PDG-PRD-98-030001-2018}. The
string tension $\kappa$ depends, in general, on the temperature,
possessing an average estimation $\sqrt{\kappa} \approx 0.405$ GeV
\cite{PRD-90-074017-2014} for cold strongly interacting matter.
The exact analytic view of the $\alpha_{s}(\mu)$ within the 1-loop
approximation is
\begin{eqnarray}
\label{eq:3b.2} \alpha_s(\mu)=(\beta_{0}t)^{-1},
\end{eqnarray}

\noindent where
$t\equiv\ln\bigl(\mu^{2}/\Lambda_{\scriptsize{\mbox{QCD}}}^{2}\bigr)$,
$\beta_{0} = (33-2n_{f})/12\pi$ is the $1$-loop $\beta$-function
coefficient, $n_{f}$ is the number of quark flavors active at the
energy scale $\mu$, i.e. are considered light $m_{q} \ll \mu$,
$m_{q}$ is the quark mass\footnote{The condition $\mu \geq
\varepsilon_{2}m_{q}$ is used for definition of the quark with
certain flavor as light one and, in the present paper, it is used
$\varepsilon_{2}=10$.}, $\Lambda_{\scriptsize{\mbox{QCD}}}$ is
non-universal scale parameter depending on the renormalization
scheme adopted, corresponding to the scale where the
perturbatively-defined coupling would diverge
\cite{PDG-PRD-98-030001-2018}.

The numerical value of $\Lambda_{\scriptsize{\mbox{QCD}}}$ depends, in particular, on $n_{f}$ and here one uses $\Lambda_{\scriptsize{\mbox{QCD}}}$ from \cite{PDG-PRD-98-030001-2018}, for a given $n_{f}$. At present-day, the convenient estimation of the $\alpha_{s}(\mu)$ is calculated within the complete 5-loop approximation \cite{PDG-PRD-98-030001-2018,JHEP-1702-090-2017}. Moreover, the running coupling constant can be defined from any physical observable perturbatively calculated \cite{grunberg}, and for each $\mu$ is obtained a $\alpha_{s}$ resulting in a specific (\ref{eq:3b.1}).

As seen from (\ref{eq:3b.2}), one must require $\mu >
\mu_{\scriptsize{\mbox{min}}} \equiv \omega
\Lambda_{\scriptsize{\mbox{QCD}}}$ to preserve the perturbative
definition validity of the coupling $\alpha_s(\mu)$. The softest
case corresponds to $\omega=1$ while more conservative and exact
estimation is given by \cite{grunberg}
\begin{eqnarray}
\label{eq:3b.3} \omega =
\exp\bigl[F_{0}(\alpha_s^{\scriptsize{\mbox{max}}})/2\beta_{0}\bigr],
\end{eqnarray}

\noindent where
$\alpha_s^{\scriptsize{\mbox{max}}}=\beta_{0}/\beta_{1}$,
$F_{0}(x)=x^{-1}+\beta_{1}/\beta_{0}\ln(\beta_{0}x)$,
$\beta_{1}=(153-9n_{f})/24\pi^{2}$ is the $2$-loop
$\beta$-function coefficient \cite{PDG-PRD-98-030001-2018}.

There are several estimation of $\mu$ based on $Y_{h}^{\scriptsize{\mbox{exp}}}$, an experimentally measurable quantity. In hadronic collisions, for instance, it is assumed $\mu=Y_{h}^{\scriptsize{\mbox{exp}}}$ at $Y_{h}^{\scriptsize{\mbox{exp}}} \equiv p_{T}^{\scriptsize{\mbox{max}}}$ \cite{PRD-86-014022-2012} or $Y_{h}^{\scriptsize{\mbox{exp}}} \equiv m_{3}$ \cite{EPJC-75-186-2015}, where $p_{T}^{\scriptsize{\mbox{max}}}$ is the transverse momentum of the leading jet, and $m_{3}$ is the invariant mass of the three jets leading in $p_{T}$.

On the other hand, in the additive quark model \cite{J.Nyiri.Int.J.Mod.Phys.A.18.2403.2003}, the $\mu$ scale can be connected with the interaction energy of the leading single $\bar{q}q$-pairs, responsible for the produced particles. The non-leading pairs, called spectators, do not contribute to the particle production \cite{EPJC-70-533-2010}. In this picture, the leading particles from the spectators carry away almost all of the collision energy, resulting in that energy has been left for the particle production is about 1/9 of the entire nucleon energy \cite{J.Nyiri.Int.J.Mod.Phys.A.18.2403.2003,EPJC-70-533-2010}. In a straight analogy, one assumes this corresponds to the scenario for which the $\bar{q}q$-pairs are subject here. Thus, a significant part of the collision energy is absorbed by the spectator $\bar{q}q$-pair, whose contribution to the elastic scattering can be neglected. Then, only part of the collision energy may be used by the (effective) leading $\bar{q}q$-pairs described by the confinement potential. One can note the relation $\mu^{2} \propto s$ is used in general scheme for running coupling in QCD \cite{Leader-book-V2-1996}.

Taking into account the above discussion, the energy scale $\mu$
may be connected with $s$ by assuming the simple relation
$\mu^{2}=\eta s$ where $0< \eta \leq 1$, which implies that $\mu$
is just a fraction of the energy involved in the elastic
scattering process\footnote{One can note that $\eta=1$ within the
approach of point-like particles used above in the Sec.
\ref{sec:4}, which corresponds to the case of interactions between
structureless fundamental constituents (fermions, bosons) of the
Standard Model at present accuracy level.}. Taking into account
the energy balance in finite-size particles collisions, one can
use $\eta=1/9$
\cite{J.Nyiri.Int.J.Mod.Phys.A.18.2403.2003,EPJC-70-533-2010},
which results in $\mu^{2}=s_{e^{+}e^{-}}$, where $s_{e^{+}e^{-}}$ is the square of the collision energy in $e^{+}e^{-}$ annihilation. The $s_{e^{+}e^{-}}$ is usually used for running coupling in QCD. Moreover, it should point out that assumption connecting the effective energy scale for hadronic collisions ($\mu$ in context of the present work) with $s_{e^{+}e^{-}}$ is not new, being widely and successfully used for decades in many studies, in particular, for the leading effect \cite{LNC-38-359-1983}, total cross section in nucleon-nucleon collisions \cite{PAN-81-508-2018} etc. It should be stressed that
$\Lambda_{\scriptsize{\mbox{QCD}}}$ is estimated only for $n_{f}
\geq 3$ \cite{PDG-PRD-98-030001-2018}. Thus, one can consider $\mu
\geq 0.96$ GeV, i.e. $\sqrt{s} \geq 2.88$ GeV based on the
perturbatively defined coupling for strong interactions, and
taking into account the condition for the lightness of the quark
with a certain flavor, as well as the relation between $\mu^{2}$
and $s$ given above. This energy range cover almost all energies
allowed for nucleon-nucleon collisions with exception of the
narrow region close to the low boundary
$\sqrt{s_{\scriptsize{\mbox{l.b.}}}}$.

The assumption performed above is analogous to the momentum
fraction $x$ carried by a scattered quark in deep inelastic
scattering. The hadron density grows as the energy increases,
since there is a change in the fractal dimension of the total
cross-section, as proposed in \cite{bc}. This can be viewed as the
parton density increasing, implying the use of very small values
for $x$. The cutoff in the parton density growth can be studied by
the Balitski--Kovchegov equation, that realizes this saturation
through pomeron fan diagrams \cite{bartels_braun}. On the other
hand, as the density grows, the distance narrows between
$\bar{q}q$ pairs and within the pairs itself.

The Helmholtz free energy can be understood here in the following way.
As the TE increases, the number of degree of freedom of
$\bar{q}q$-pairs rise. Thus, the internal energy is given mostly
by pairs of particles in the non-confinement regime, i.e. these
pairs approach the asymptotic freedom. Then, the entropy term
may dominate over the confinement potential and this information
should be taken into account in the Helmholtz free energy.
However, it is expected this situation may be achieved only near
the Hagedorn temperature $T_{c}$. On the other hand, when the
entropy diminishes the number of degree of freedom also diminish
turning the confinement potential the main energy source. This explanation is the same in the case of the BKT-phase transition in terms of the transition temperature \cite{V.L.Berezinskii.Zh.Eksp.Teor.Fiz.59.907.1970,J.M.Kosterlitz.D.J.Thouless.J.Phys.C6.1181.1973}. Below the transition temperature, the potential energy dominates, preventing the emergence of a single vortex. Otherwise, the entropy is favored, turning possible the presence of a single vortex state.

\subsection{Confinement potential in $b$-space and normalization procedure}

The confinement potential is of short-range in contrast with the
Coulomb one and, by reason of the uncertainty principle, the
quantity $\mu_{\scriptsize{\mbox{min}}}$ allows the unambiguous
estimation of the linear scale $r_{\scriptsize{\mbox{max}}} \sim
\mu_{\scriptsize{\mbox{min}}}^{-1}$, up to which the confinement
potential can be calculated with help of (\ref{eq:3b.1}). One can
expect $r_{\scriptsize{\mbox{max}}} \sim R_{h}$, depending on the
approach for $\mu_{\scriptsize{\mbox{min}}}$ and on the values of
the $\Lambda_{\scriptsize{\mbox{QCD}}}$ at given $n_{f}$
\cite{PDG-PRD-98-030001-2018}, where $R_{h}$ is the hadron radius.
This upper cutoff for $r$ tames the divergence of the confinement
potential (\ref{eq:3b.1}). In general, one should considers $b\leq
b_{\scriptsize{\mbox{max}}} \sim r_{\scriptsize{\mbox{max}}}$ for
the incoming particles interacting by strong force with each
other. Table \ref{tab:4.1} shows the values for $b_{\scriptsize{\mbox{max}}}=\mu_{\scriptsize{\mbox{min}}}^{-1}$ calculated for various numbers of light quark flavor and schemes, aiming the definition of the low boundary for the domain on $\mu$ in which the perturbative definition of the coupling $\alpha_s(\mu)$ is valid. As seen, $b_{\scriptsize{\mbox{max}}}$ is significantly smaller within a conservative scheme for $\mu_{\scriptsize{\mbox{min}}}$ than that for softest one at any fixed $n_{f}$ and values of $b_{\scriptsize{\mbox{max}}}$ are in
the range from about 0.59 (0.37) fm at $\sqrt{s}=2.88$ GeV to the
$\simeq 2.22$ (1.44) fm at the nominal LHC energy $\sqrt{s}=14$
TeV for the softest (conservative) restriction on the
$\mu_{\scriptsize{\mbox{min}}}$. As expected
$b_{\scriptsize{\mbox{max}}}(s)$ is constant at fixed $n_{f}$ with
sharply increasing at growth of $n_{f}$. The step magnitude
increases with the onset of the influence of heavier flavors,
being largest for the transition from $n_{f}=5$ to $n_{f}=6$. On
the other hand, a smaller space scale $r \to 0$ inside the hadron
can be probed through more central collisions, with $b \to 0$.
Within the general framework of the paper, the relation
$b_{\scriptsize{\mbox{min}}} \sim \mu^{-1}=(\eta s)^{-1/2}$ is
used for a rough estimation of the lower boundary for the impact
parameter at a given $s$. Therefore, one can assume
$r=\varepsilon_{3}b$, where $\varepsilon_{3} \leq 1$, and the
confinement potential in the impact parameter space can be
rewritten as
\begin{eqnarray} \label{eq:3b.4} V_{\scriptsize{\mbox{c}}}(\mu,r) = V_{\scriptsize{\mbox{c}}}(s,b) =
-4\alpha_s(\eta s)/3\varepsilon_{3}b+\kappa \varepsilon_{3} b.
\end{eqnarray}
The potential $V_{\scriptsize{\mbox{c}}}(\mu,r)$ and
$V_{\scriptsize{\mbox{c}}}(s,b)$ coincide exactly in whole domain
$(\mu \geq \mu_{\scriptsize{\mbox{min}}};r \leq
r_{\scriptsize{\mbox{max}}}$), due to exact (linear)
interrelations between the corresponding terms in the parameter
pairs $(\mu,r)$ and $(s,b)$. Note the Coulomb-like term in $V_{\scriptsize{\mbox{c}}}(s,b)$ behaves as $V_{\scriptsize{\mbox{C}}}(s,b)\approx-ab^{-1}$ for a sufficient high fixed-$s$. However, in $V_{\scriptsize{\mbox{c}}}(s,b)$ the information about $s$ is embodied in the running coupling.

Taking into account the general
properties of hadronic collisions discussed above, for the sake of
simplicity, one uses $\varepsilon_{3}=1.0$ unless otherwise
specifically indicated. As seen the confinement potential
(\ref{eq:3b.4}) is null at
\begin{eqnarray}
\label{eq:3b.5} b_{0}=\sqrt{4\alpha_s(\eta s)/3\kappa} \propto
\sqrt{\alpha_s(\eta s)},
\end{eqnarray}

\noindent where the allowable ranges are taken into account for
the linear scales $r$ and $b$, i.e. $r, b > 0$. Table \ref{tab:4.2} shows the values of $b_{0}$ calculated for various $n_{f}$ within 1- and 5-loop approximation for $\alpha_{s}(\mu)$
	at two collision energies considered in the present work. Transition from the intermediate energy $\sqrt{s}=52.8$ GeV to the high one $\sqrt{s}=14$ TeV significantly reduces $b_{0}$ from $\sim 0.21-0.24$ fm down to $\sim 0.14-0.16$ fm at fixed $n_{f}$. The increase in the number of light quark flavors results in the growth of $b_{0}$ for some collision energies, whereas the use of higher-order approximation for $\alpha_{s}(\mu)$ provides smaller values of $b_{0}$ at fixed $n_{f}$ and $s$. It is
interesting to note that it can be shown that $b_{0} \gtrsim
b_{\scriptsize{\mbox{max}}}$ at $\sqrt{s} \approx 2.88$ GeV for
$\omega$ defined by (\ref{eq:3b.3}) and 5-loop approximation, i.e.
in this case $\forall\,b: V_{\scriptsize{\mbox{c}}}(s,b) < 0$, in
the very narrow energy range close to the lowest allowed value of
$s$.

Thus the detailed analysis shows that the characteristic linear
scales in the impact parameter space --
$b_{\scriptsize{\mbox{min}}}$, $b_{0}$ and
$b_{\scriptsize{\mbox{max}}}$ -- are $s$-dependent. There is also
a dependence on the scheme for the estimation of
$\mu_{\scriptsize{\mbox{min}}}$ for the
$b_{\scriptsize{\mbox{max}}}$ (Table \ref{tab:4.1}) as well as
there is a relies on the number of loops for $\alpha_{s}(\eta s)$
approximation for $b_{0}$ (Table \ref{tab:4.2}).

\begin{table}
\caption{Maximum values of the impact parameter for perturbative
approach ($b_{\scriptsize{\mbox{max}}}$, fm)} \label{tab:4.1}
\begin{center}
\begin{tabular}{lcccc}
\hline \multicolumn{1}{l}{Scheme for
$\mu_{\scriptsize{\mbox{min}}}$} & \multicolumn{4}{c}{$n_{f}$}
\rule{0pt}{10pt}\\\cline{2-5}
 & 3 & 4 & 5 & 6  \rule{0pt}{10pt}\\
\hline
softest ($\omega=1$)        & $0.59 \pm 0.03$   & $0.68 \pm 0.04$   & $0.94 \pm 0.06$ & $2.22 \pm 0.15$ \rule{0pt}{10pt}\\
conservative ($\omega > 1$) & $0.365 \pm 0.019$ & $0.418 \pm 0.023$ & $0.59 \pm 0.04$ & $1.44 \pm 0.10$ \rule{0pt}{10pt}\\
\hline
\end{tabular}
\end{center}
\end{table}

\begin{table}
\caption{Values of $b_{0}$ (fm) at two collision energies}
\label{tab:4.2}
\begin{center}
\begin{tabular}{lcccc}
\hline \multicolumn{1}{l}{Approximation} &
\multicolumn{4}{c}{$n_{f}$} \rule{0pt}{10pt}\\\cline{2-5}
 order for $\alpha_{s}(\mu)$ & 3 & 4 & 5 & 6  \rule{0pt}{10pt}\\
\hline
\multicolumn{5}{c}{$\sqrt{s}=52.8$ GeV}\rule{0pt}{10pt}\\
\hline
1-loop & $0.2359 \pm 0.0015$ & $0.2413 \pm 0.0016$ &  & \rule{0pt}{10pt}\\
5-loop & $0.2134 \pm 0.0012$ & $0.2199 \pm 0.0013$ &  & \rule{0pt}{10pt}\\
\hline
\multicolumn{5}{c}{$\sqrt{s}=14$ TeV}\rule{0pt}{10pt}\\
\hline
1-loop & $0.1521 \pm 0.0004$ & $0.1570 \pm 0.0004$ & $0.1610 \pm 0.0005$ & $0.1617 \pm 0.0005$ \rule{0pt}{10pt}\\
5-loop & $0.1433 \pm 0.0004$ & $0.1486 \pm 0.0004$ & $0.1534 \pm 0.0005$ & $0.1558 \pm 0.0005$ \rule{0pt}{10pt}\\
\hline
\end{tabular}
\end{center}
\end{table}

\begin{figure}[ht]
\centering
\includegraphics[width=16.0cm,height=16.0cm]{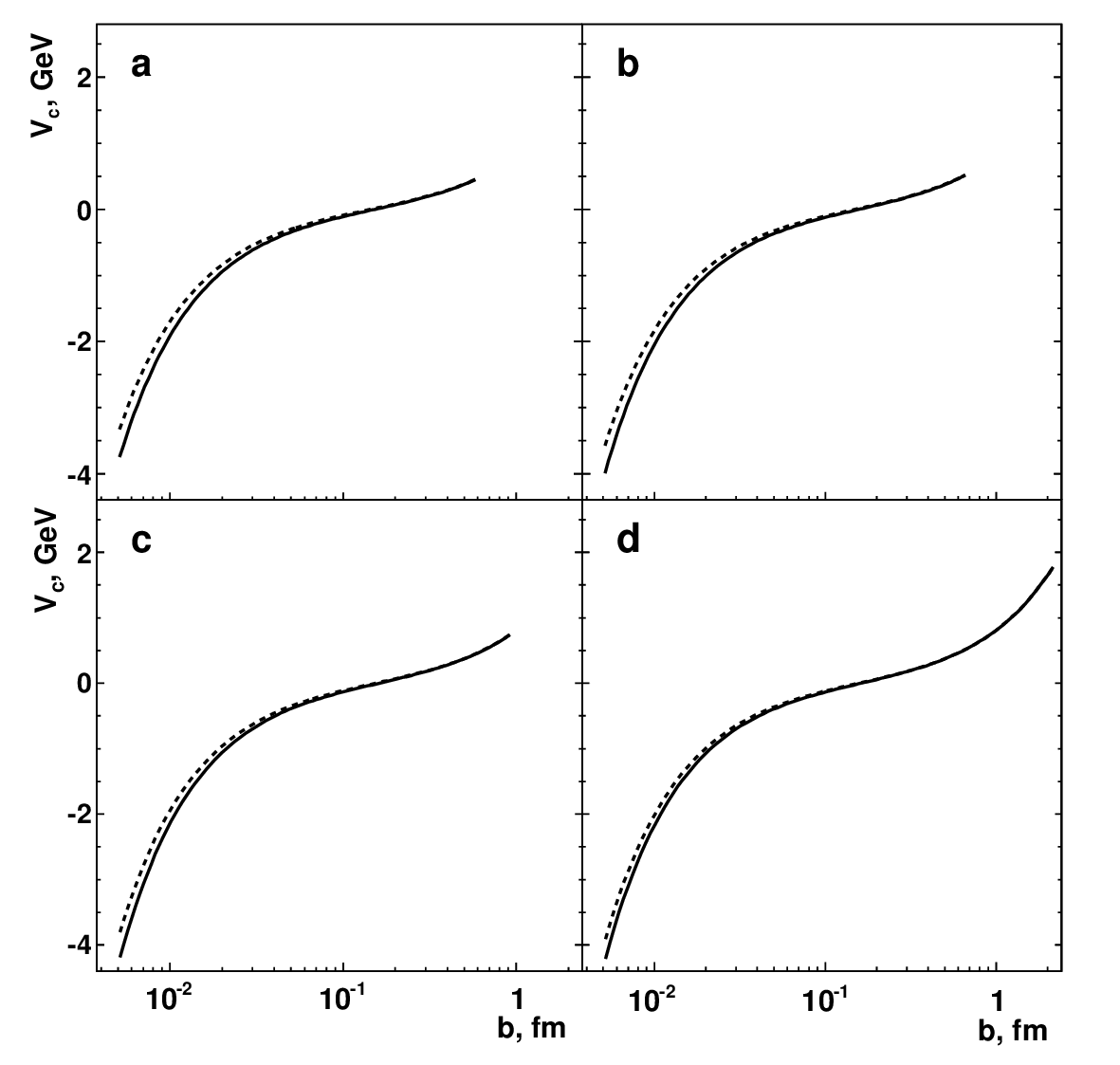}
\caption{\label{fig:4}Dependence of the confinement potential at $\sqrt{s}=14$ TeV and $n_{f}=3$ (a), $n_{f}=4$ (b), $n_{f}=5$ (c) and $n_{f}=6$ (d) with various approximations for $\alpha_{s}(\mu)$: solid line is for the 1-loop solution (\ref{eq:3b.2}) and dashed curve corresponds to the 5-loop approximation \cite{PDG-PRD-98-030001-2018}. The softest scheme for $\mu_{\scriptsize{\mbox{min}}}$ is used without loss of generality.}
\end{figure}

Contrary to the Coulomb potential, the confinement allows a glance
at the hadron internal arrangement, revealing its importance for
the correct description of the elastic scattering, even in this
naive potential approach. The Fig. \ref{fig:4} shows the
$b$-dependence of the confinement potential (\ref{eq:3b.4}) for
fixed $\sqrt{s}=14$ TeV and various numbers of the light flavors
$n_{f}$ for 1- and 5-loop approximation for $\alpha_{s}(\mu)$. The
value $\omega=1$ is used for definition of the
$\mu_{\scriptsize{\mbox{min}}}$ and, consequently, the largest
value of the upper cutoff for $b$.

\begin{figure}[ht]
\centering
\includegraphics[width=16.0cm,height=16.0cm]{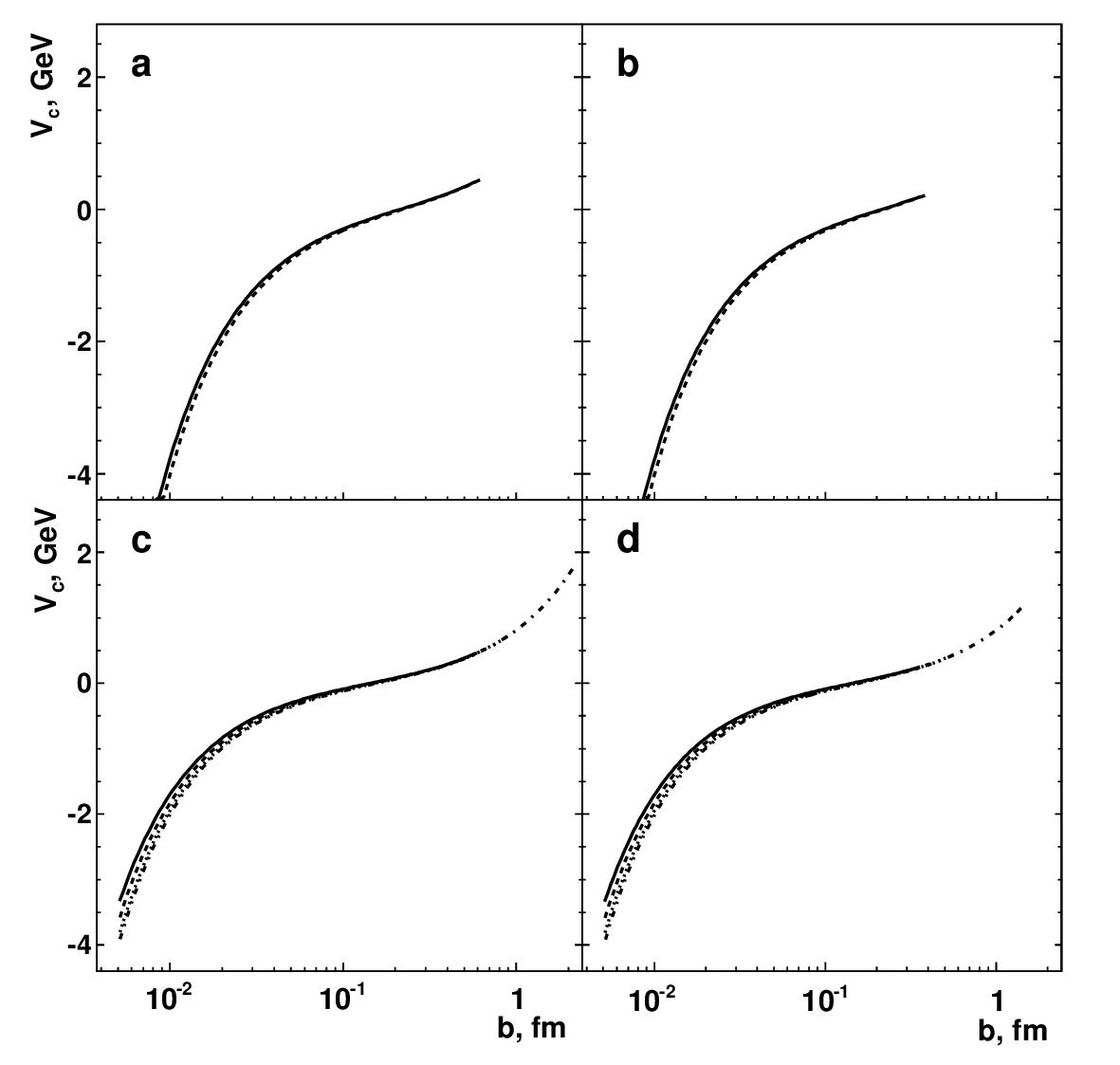}
\caption{\label{fig:5}Dependence of the confinement potential on impact parameter at $\sqrt{s}=52.8$ GeV (a, b) and $\sqrt{s}=14$ TeV (c, d) for various $n_{f}$: solid line is for $n_{f}=3$, dashed one -- for $n_{f}=4$, dotted curve corresponds to the $n_{f}=5$ and dot-dashed one -- to the $n_{f}=6$. The left column (a, c) shows results for $\omega=1$ and curves for conservative estimation (\ref{eq:3b.3}) are in the right column (b, d).}
\end{figure}

The 5-loop approximation for $\alpha_{s}(\mu)$ provides slightly
larger values of $V_{\scriptsize{\mbox{c}}}(s,b)$ than that for
1-loop exact solution (\ref{eq:3b.2}) only in the range of small
values $b \lesssim 0.03$ fm (not shown here). The consistent
transition from Fig. \ref{fig:4}a to Fig. \ref{fig:4}d shows the
weakening of the difference between the two curves with the growth
of $n_{f}$. Thus, in general the approximation order for
$\alpha_{s}(\mu)$ influences weakly on the
$V_{\scriptsize{\mbox{c}}}(s,b)$ in the whole range of $b$ for any
$n_{f}$ considered, and the results are stable with respect to the
scheme of calculation for $\alpha_{s}(\mu)$. By definition, the
modern 5-loop approximation can be used for $\alpha_{s}(\mu)$
below, unless otherwise indicated. At the intermediate energy
$\sqrt{s}=52.8$ GeV, $V_{\scriptsize{\mbox{c}}}(s,b)$ does not
depend on the number of light flavors (Figs. \ref{fig:5}a, b). The
confinement potential dependence on $n_{f}$ manifests itself only
in the high energy domain, for instance at $\sqrt{s}=14$ TeV, in
which the wide set of the values of $n_{f}$ is available (Figs.
\ref{fig:5}c, d). In the last case, the growth of $n_{f}$ has
provided some decrease in $V_{\scriptsize{\mbox{c}}}(s,b)$ for
small $b \lesssim 0.03$ fm as expected, expanding the confinement
potential for larger impact parameter values $b \gtrsim R_{h}$ due
to the decrease of $\Lambda_{\scriptsize{\mbox{QCD}}}$.

\begin{figure}[ht]
\centering
\includegraphics[width=16.0cm,height=16.0cm]{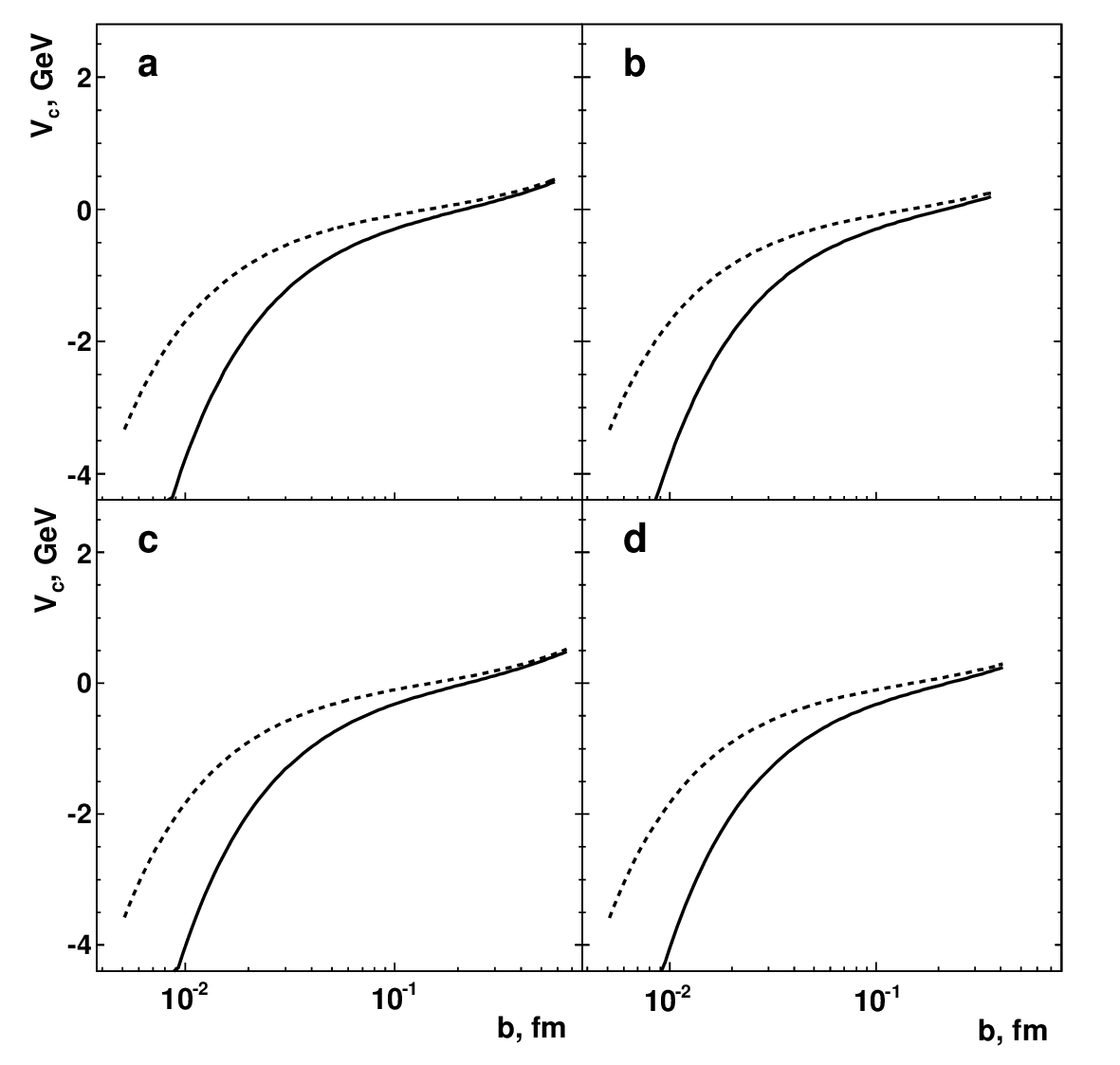}
\caption{\label{fig:6}Dependence of the confinement potential on impact parameter at $n_{f}=3$ (a, b) and $n_{f}=4$ (c, d) for two collision energies: solid line is for $\sqrt{s}=52.8$ GeV and dashed one -- for $\sqrt{s}=14$ TeV. The left column (a, c) shows results for $\omega=1$ and curves for conservative estimation (\ref{eq:3b.3}) are in the right column (b, d).}
\end{figure}

The modification in the scheme to estimate of
$\mu_{\scriptsize{\mbox{min}}}$ does not influence on the
functional behavior of the $V_{\scriptsize{\mbox{c}}}(s,b)$, for
both the intermediate (Figs. \ref{fig:5}a, b) and the high energy
(Figs. \ref{fig:5}c, d) considered here. The transition from
$\omega=1$ (Figs. \ref{fig:5}a  c) to the conservative estimation
of this parameter (Figs. \ref{fig:5}b, d), leads to the decrease
of the high boundaries for linear scales $r$ and $b$. The Fig.
\ref{fig:6} shows the evolution of the
$V_{\scriptsize{\mbox{c}}}(s,b)$  considering the collision energy
growth for fixed $n_{f}=3$ (a, b) and $n_{f}=4$ (c, d) for two
different approaches for $\mu_{\scriptsize{\mbox{min}}}$. As seen
before, $V_{\scriptsize{\mbox{c}}}(s,b)$ is larger for
$\sqrt{s}=14$ TeV than that for $\sqrt{s}=52.8$ GeV at
corresponding values of $b$, for any number of light flavors
$n_{f}$ and scheme for the $\omega$-parameter calculation.
Furthermore, the difference between the two curves increases as
$b$ decreases. The behavior of $V_{\scriptsize{\mbox{c}}}(s,b)$ in
Fig. \ref{fig:6} is explained by the smooth decreasing of
$\alpha_{s}(\mu)$ with the growth of $\mu$
\cite{PDG-PRD-98-030001-2018}, i.e. with the collision energy
growth due to the relation used here.

In general, the main features of the confinement potential shown
in Figs. \ref{fig:4} -- \ref{fig:6} are driven by contributions
coming from different terms in (\ref{eq:3b.1}) or, consequently,
(\ref{eq:3b.4}) for several ranges of the impact parameter values.
The $V_{\scriptsize{\mbox{c}}}(s,b)$ is sensitive for changes in
$n_{f}$, $s$ and mostly for small $b$, since the main contribution
in this range comes from the first (short-range) term in
(\ref{eq:3b.4}) containing $\alpha_{s}(\mu)$ that depends, in
turn, on $n_{f}$ and $s$. The influence of the first term
decreases as $b$ grows as well as the contribution of the second
(long-range) term becomes dominant in (\ref{eq:3b.4}). This term
depends on string tension only and, consequently,
$V_{\scriptsize{\mbox{c}}}(s,b)$ it is not sensitive for $n_{f}$
and changes weakly with $s$, for relatively large $b > 0.1$ fm.
Here, changes of $n_{f}$ and / or $s$ provides different values
for the up boundary $b_{\scriptsize{\mbox{max}}}$ for $b$--range
considered perturbatively.

It is necessary to normalize the potential (\ref{eq:3b.4}) to obey
the unitarity condition. As well-known, the second term in
(\ref{eq:3b.1}) as well as (\ref{eq:3b.4}) provides the main
difference between the confinement potential and the Coulomb one,
namely, the positive values and the quasi-linear growth of the
$V_{\scriptsize{\mbox{c}}}(s,b)$ at large $b$, i.e. $r$ (Figs.
\ref{fig:4} -- \ref{fig:6}). If one considers the equation
(\ref{eq:3.3}) and taking into account the appropriate values of
$\gamma$, then the confinement potential has a constant sign
within the $b$-range under consideration. However, in general, the
potential $V(s,b)$ may change its sign within the kinematic domain
studied and this feature can lead to the discontinuity for
$\tilde{V}(s,b)$, if extremum (maximum) value of the $V(s,b)$ is
used as the scale factor. The analysis performed shows that using
the maximum for the absolute value of the potential, then
$|V(s,b)|_{\scriptsize{\mbox{max}}}$ avoids the possible
discontinuity in the behavior of the $\tilde{V}(s,b)$, in the case
of sign changing of $V(s,b)$. Therefore, here the following
relation is used
\begin{eqnarray}
\label{eq:3b.6}\tilde{V}_{\scriptsize{\mbox{c}}}(s,b)=|V_{\scriptsize{\mbox{c}}}(s,b)|/
|V_{\scriptsize{\mbox{c}}}(s,b)|_{\scriptsize{\mbox{max}}}.
\end{eqnarray}

\noindent where $|V_{\scriptsize{\mbox{c}}}(s,b)|$ is the absolute value of the confinement potential $V_{\scriptsize{\mbox{c}}}(s,b)$, if the confinement potential changes its sign within the $b$-range under discussion. In this case, the $|V_{\scriptsize{\mbox{c}}}(s,b)|_{\scriptsize{\mbox{max}}}$ can be reached at low or high boundary of $b$ (Figs. \ref{fig:4} -- \ref{fig:6}). Without loss of generality, the range $b_{\scriptsize{\mbox{min}}} \leq b \leq b_{\scriptsize{\mbox{max}}}$ is studied below, where $V_{\scriptsize{\mbox{c}}}(s,b_{\scriptsize{\mbox{min}}}) < 0$ and $b_{\scriptsize{\mbox{max}}}$ is controlled by $r_{\scriptsize{\mbox{max}}}$. Thus, the normalized confinement potential, on the impact parameter space, can be written as
\begin{eqnarray}
\tilde{V}_{\scriptsize{\mbox{c}}}(s,b)=
\frac{b}{b_{\scriptsize{\mbox{n}}}}\biggl|\frac{1-[4\alpha_s(\eta
s)/3\kappa]b^{-2}}{1-[4\alpha_s(\eta s)/3\kappa]
b_{\scriptsize{\mbox{n}}}^{-2}}\biggr| =
\frac{b}{b_{\scriptsize{\mbox{n}}}}\biggl|\frac{1-(b_{0}/b)^{2}}{1-(b_{0}/b_{\scriptsize{\mbox{n}}})^{2}}\biggr|,
\label{eq:3b.7.new}
\end{eqnarray}
\noindent where
\begin{eqnarray}
b_{\scriptsize{\mbox{n}}}=\left\{
\begin{array}{lcl}
b_{\scriptsize{\mbox{min}}}, & \mbox{if} &
|V_{\scriptsize{\mbox{c}}}(s,b_{\scriptsize{\mbox{min}}})|
> V_{\scriptsize{\mbox{c}}}(s,b_{\scriptsize{\mbox{max}}}); \vspace{0.2cm}\\
b_{\scriptsize{\mbox{max}}}, & \mbox{if} & |V_{\scriptsize{\mbox{c}}}(s,b_{\scriptsize{\mbox{min}}})| \leq V_{\scriptsize{\mbox{c}}}(s,b_{\scriptsize{\mbox{max}}}). \\
\end{array}\right.\label{eq:3b.7}
\end{eqnarray}

The confinement potential is still assumed as featured by finite
negative value for $b_{\scriptsize{\mbox{min}}}$\footnote{One can
note that there is no limit for $b_{\scriptsize{\mbox{min}}}$,
since it can be $\mu \to \infty$ in general. The present
experimental restriction on the size of fundamental constituents
of the Standard Model can be suggested as the estimation of the
low boundary ($b_{\scriptsize{\mbox{l.b.}}}$) of the
$b_{\scriptsize{\mbox{min}}}$ in the relation (\ref{eq:3b.7}):
$b_{\scriptsize{\mbox{min}}} \geq b_{\scriptsize{\mbox{l.b.}}}
\sim 2 \times 10^{-4}$ fm at $\mu_{\scriptsize{\mbox{max}}} \sim
1$ TeV \cite{PDG-PRD-98-030001-2018}.}.

As seen above,
$V_{\scriptsize{\mbox{c}}}(s,b_{\scriptsize{\mbox{max}}}) < 2$ GeV
for any loop approximation (Fig. \ref{fig:4}), scheme for
$\mu_{\scriptsize{\mbox{min}}}$ estimation, and $n_{f}$ values
(Fig. \ref{fig:5}). Consequently, the condition
$|V_{\scriptsize{\mbox{c}}}(s,b)|_{\scriptsize{\mbox{max}}} =
V_{\scriptsize{\mbox{c}}}(s,b_{\scriptsize{\mbox{max}}})$ is valid
up to $b_{\scriptsize{\mbox{min}}} \gtrsim 10^{-2}$ fm. Therefore,
the lower relation in (\ref{eq:3b.7}) is, in general, applicable,
while the upper equation in (\ref{eq:3b.7}) is valid only for
processes that probe the inner structure of a hadron down to the
very small linear scales.

\subsection{Inelasticity and TE for the strong interaction}

\begin{figure}[ht]
\centering
\includegraphics[width=16.0cm,height=16.0cm]{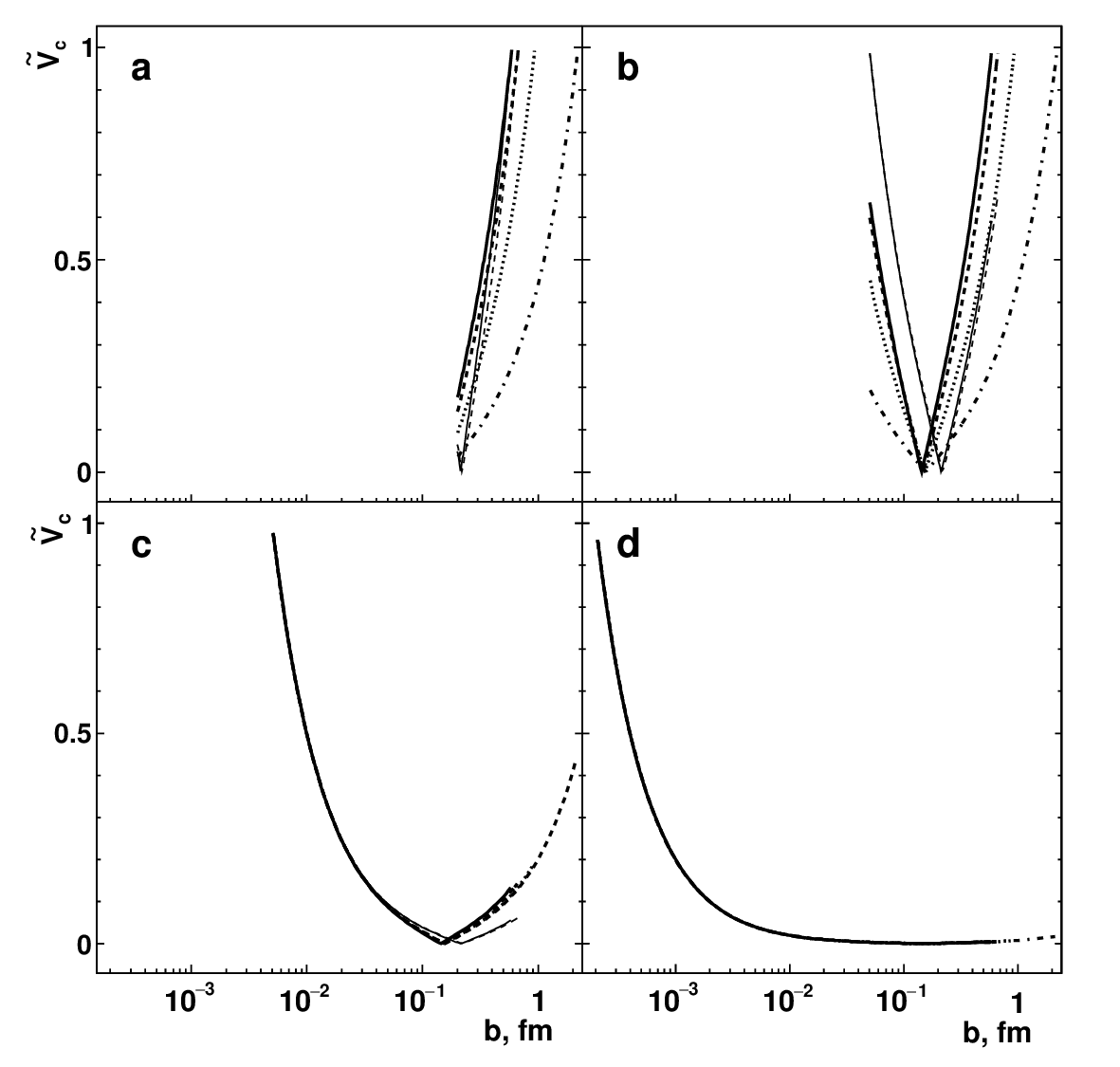}
\caption{\label{fig:after6}Dependence of the normalized confinement potential on impact parameter with the following values of the low boundary for $b$--range under studied: $0.2$ fm (a), $0.05$ fm (b), $5 \times 10^{-3}$ fm (c) and $2 \times 10^{-4}$ fm (d). Are considered two collision energies $\sqrt{s}=52.8$ GeV (thin lines) and $\sqrt{s}=14$ TeV (thick curves). The $\tilde{V}_{\scriptsize{\mbox{c}}}(s,b)$ are shown for $\omega=1$ and solid lines are for $n_{f}=3$, dashed ones -- for $n_{f}=4$, dotted curves correspond to the $n_{f}=5$ and dot-dashed ones -- to the $n_{f}=6$.}
\end{figure}

Results are shown in Fig. \ref{fig:after6} for detailed analysis of the dependence of $\tilde{V}_{\scriptsize{\mbox{c}}}(s,b)$ on the impact parameter for several $s$, $n_{f}$ and ranges $b \in[b_{\scriptsize{\mbox{min}}},b_{\scriptsize{\mbox{max}}}]$. The softest scheme for $\mu_{\scriptsize{\mbox{min}}}$ is used, without loss of generality. In Fig. \ref{fig:after6}a the relations $b_{\scriptsize{\mbox{n}}}=b_{\scriptsize{\mbox{max}}} \gg b_{0}$ are valid. In this case
\begin{eqnarray*}
\tilde{V}_{\scriptsize{\mbox{c}}}(s,b) \approx
b\bigl[1-(b_{0}/b)^{2}\bigr] / b_{\scriptsize{\mbox{max}}}.
\end{eqnarray*}

Here the $n_{f}$-- and $s$--dependencies survive in $\tilde{V}_{\scriptsize{\mbox{c}}}(s,b)$ due to $b_{0}$ and $b_{\scriptsize{\mbox{max}}}$. These dependencies are seen most clearly in Fig. \ref{fig:after6}b. The minimum of $\tilde{V}_{\scriptsize{\mbox{c}}}(s,b)$ goes to the smaller $b$ with the increase of the dip for larger $s$ and fixed $n_{f}$, in accordance with the dependence $\alpha_s(\eta s)$. The relations $b_{\scriptsize{\mbox{n}}}=b_{\scriptsize{\mbox{min}}} \ll b_{0}$ are valid in Figs. \ref{fig:after6}c, d. Then
\begin{eqnarray*}
\tilde{V}_{\scriptsize{\mbox{c}}}(s,b) \approx
bb_{\scriptsize{\mbox{min}}}\bigl[1-(b_{0}/b)^{2}\bigr] /
b_{0}^{2}.
\end{eqnarray*}

This equation allows two asymptotic cases: (i)
$\bigl.\tilde{V}_{\scriptsize{\mbox{c}}}(s,b)\bigr|_{b \to
b_{\scriptsize{\mbox{min}}} \ll b_{0}} \to
b_{\scriptsize{\mbox{min}}}/b$ and (ii)
$\bigl.\tilde{V}_{\scriptsize{\mbox{c}}}(s,b)\bigr|_{b \to
b_{\scriptsize{\mbox{max}}} \gg b_{0}} \to
bb_{\scriptsize{\mbox{min}}}/b_{0}^{2}$. As can be seen above,
there are no $n_{f}$-- and $s$--dependencies of the normalized
confinement potential for values of $b$ close to the down boundary
$b_{\scriptsize{\mbox{min}}}$ of the considered range. At large $b
\to b_{\scriptsize{\mbox{max}}}$, the energy and
$n_{f}$--dependencies display itself due to $b_{0}$ but these
dependencies are (very) weak because of (very) small
$b_{\scriptsize{\mbox{min}}}$. The parameter
$b_{\scriptsize{\mbox{max}}}$ is most sensitive for changes of
$n_{f}$ and / or $s$. Figs. \ref{fig:after6}c, d confirm the
results for the asymptotic behavior of
$\tilde{V}_{\scriptsize{\mbox{c}}}(s,b)$ in the cases (i) and
(ii).

Therefore, the general conclusions follow from relations
(\ref{eq:3b.7.new}) in the domain of validity of the condition
$b_{\scriptsize{\mbox{min}}} \ll b_{0}$. The normalized
confinement potential and corresponding inelastic overlap function
$G_{\scriptsize{\mbox{inel}}}^{\,\scriptsize{\mbox{c}}}(s,b)$ are
weakly sensitive on changes of the $n_{f}$ and $s$, and the energy
dependence of the TE $S_{T}^{\,\scriptsize{\mbox{c}}}$ is driven
by $k$.

Based on the Fig. \ref{fig:after6},  $b_{\scriptsize{\mbox{min}}}=0.05$ fm is used in order to show clearly the $n_{f}$-- and $s$--dependencies of the inelastic overlap function for confinement potential.

\begin{figure}[ht]
\centering
\includegraphics[width=16.0cm,height=16.0cm]{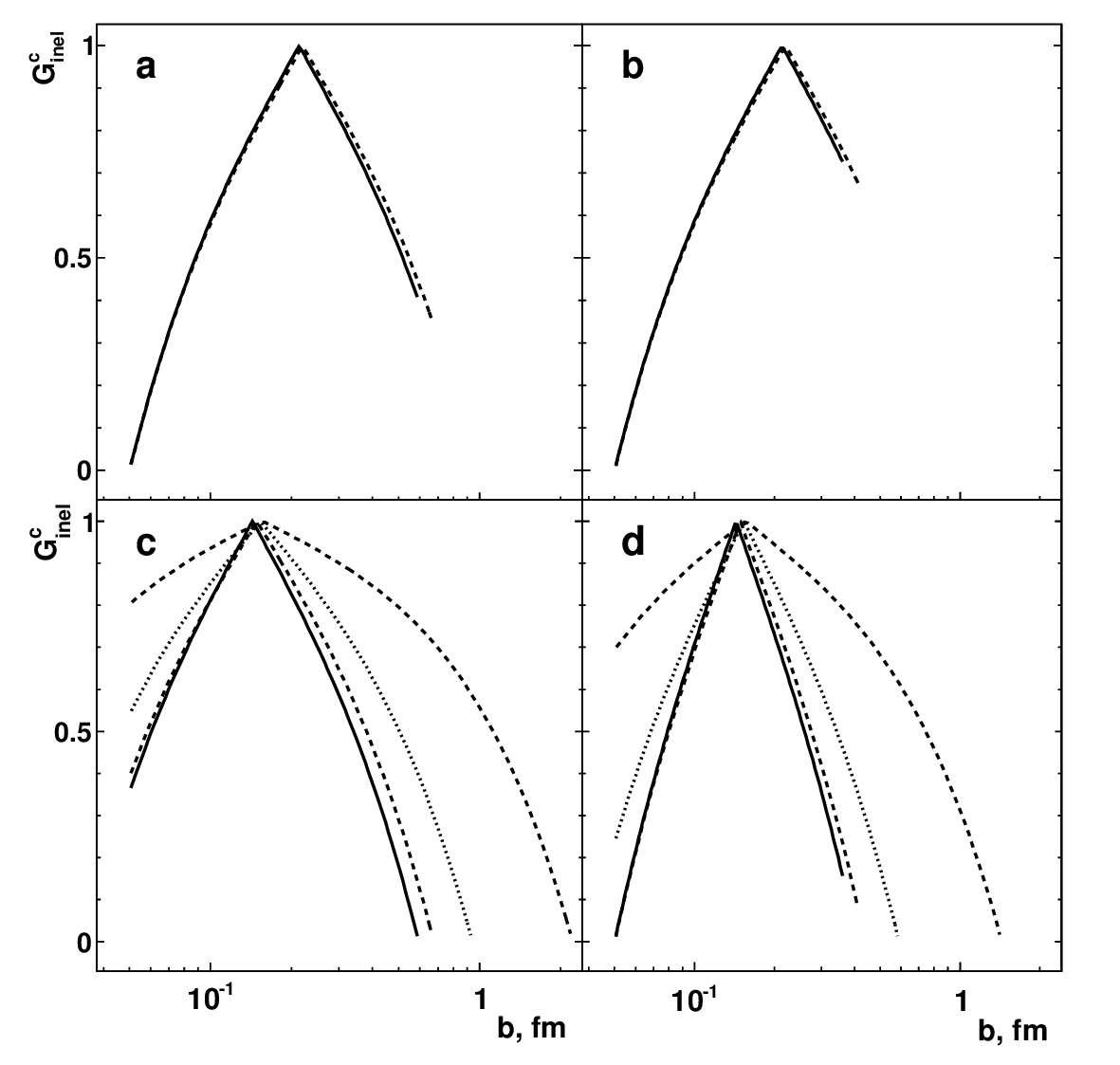}
\caption{\label{fig:7}Behavior of
$G_{\scriptsize{\mbox{inel}}}^{\,\scriptsize{\mbox{c}}}(s,b)$
using the confinement potential at $b_{\scriptsize{\mbox{min}}}=0.05$ fm, $\sqrt{s}=52.8$ GeV (a, b) and $\sqrt{s}=14$ TeV (c, d) for various $n_{f}$: solid line is for $n_{f}=3$, dashed one -- for $n_{f}=4$, dotted curve corresponds to the $n_{f}=5$ and dot-dashed one -- to the $n_{f}=6$. The left column (a, c) shows results for $\omega=1$ and curves for conservative estimation (\ref{eq:3b.3}) are in the right column (b, d).}
\end{figure}

The Fig. \ref{fig:7} shows the dependence of the inelastic overlap
function for the confinement potential
$G_{\scriptsize{\mbox{inel}}}^{\,\scriptsize{\mbox{c}}}(s,b)$ on
$b$ for several $n_{f}$ and approaches for $\omega$ for two
collision energies $\sqrt{s}=52.8$ GeV (a, b) and 14 TeV (c, d).
The scheme for estimation of $\omega$ does not influence on the
$G_{\scriptsize{\mbox{inel}}}^{\,\scriptsize{\mbox{c}}}(s,b)$ for
intermediate energies (Fig. \ref{fig:7}a, b). However, the
situation changes at $\sqrt{s}=14$ TeV (Fig. \ref{fig:7}c, d): the
conservative estimation for $\omega$ leads to smaller
$G_{\scriptsize{\mbox{inel}}}^{\,\scriptsize{\mbox{c}}}(s,b)$ at
$b < b_{0}$, in comparison with the case for $\omega=1$, and the
influence is weaker for larger $n_{f}$. The influence of $n_{f}$
on the view of
$G_{\scriptsize{\mbox{inel}}}^{\,\scriptsize{\mbox{c}}}(s,b)$ is
negligible at $\sqrt{s}=52.8$ GeV (Fig. \ref{fig:7}a, b), and the
growth of the number of light flavors leads to the increase of
$G_{\scriptsize{\mbox{inel}}}^{\,\scriptsize{\mbox{c}}}(s,b)$ at
fixed $b$ for $\sqrt{s}=14$ TeV (Fig. \ref{fig:7}c, d), especially
for large $n_{f}=5$ and 6. For the confinement potential, the
black disk regime
$G_{\scriptsize{\mbox{inel}}}^{\,\scriptsize{\mbox{c}}}(s,b)
\approx 1$ is reached at $b_{0}$ and in the region close to this
inflection point of the $V_{\scriptsize{\mbox{c}}}(s,b)$. As
discussed above, this region expands as $n_{f}$ grows, especially
for the largest $n_{f}=6$. Such behavior agrees with the
expectation for qualitative expansion of the region with high
absorption in the nucleon-nucleon collisions at higher energies.
On the other hand, the general feature of the
$G_{\scriptsize{\mbox{inel}}}^{\,\scriptsize{\mbox{c}}}(s,b)$ in
Fig. \ref{fig:7} is the more transparent (gray) regions for both
the small ($b \ll b_{0}$) and the large ($b \gg b_{0}$) impact
parameters.

\begin{figure}[ht]
\centering
\includegraphics[width=16.0cm,height=16.0cm]{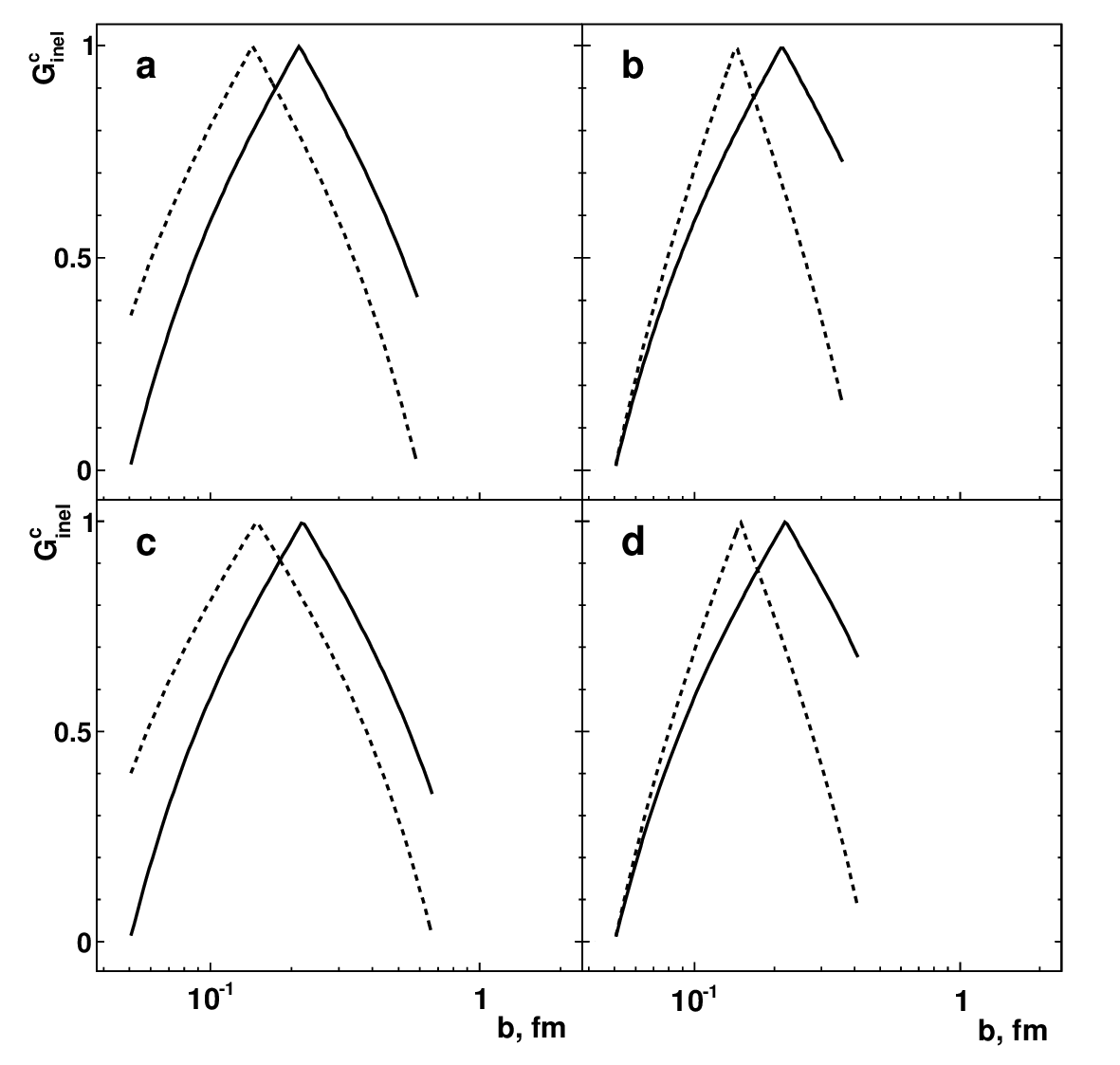}
\caption{\label{fig:8}Behavior of
$G_{\scriptsize{\mbox{inel}}}^{\,\scriptsize{\mbox{c}}}(s,b)$
using the confinement potential at $b_{\scriptsize{\mbox{min}}}=0.05$ fm, $n_{f}=3$ (a, b) and $n_{f}=4$ (c, d) for two collision energies: solid line is for $\sqrt{s}=52.8$ GeV and dashed one -- for $\sqrt{s}=14$ TeV. The left column (a, c) shows results for $\omega=1$ and curves for conservative estimation (\ref{eq:3b.3}) are in the right column (b, d).}
\end{figure}

The Fig. \ref{fig:8} shows $G_{\scriptsize{\mbox{inel}}}^{\,\scriptsize{\mbox{c}}}(s,b)$,
depending on $b$ for several $\sqrt{s}$ and approaches for $\omega$, considering two different numbers of light flavors $n_{f}=3$ (a, b) and 4 (c, d). The behavior of $G_{\scriptsize{\mbox{inel}}}^{\,\scriptsize{\mbox{c}}}(s,b)$ at $\sqrt{s}=14$ TeV, depending on the scheme for the estimation of $\omega$, leads to different relations between the two inelastic overlap functions at $n_{f}=3$ in Figs. \ref{fig:8}a and \ref{fig:8}b, at $n_{f}=4$ in Figs. \ref{fig:8}c and \ref{fig:8}d. The maximum of $G_{\scriptsize{\mbox{inel}}}^{\,\scriptsize{\mbox{c}}}(s,b)$ tends to the smaller $b \simeq 0.15$ fm as $\sqrt{s}$ increase. At $\omega=1$, the $G_{\scriptsize{\mbox{inel}}}^{\,\scriptsize{\mbox{c}}}(s,b)$ is significantly larger at $\sqrt{s}=14$ TeV than for $\sqrt{s}=52.8$ GeV at $b \lesssim 0.15$ fm, and vice versa at larger $b \gtrsim 0.3$ fm, for both the $n_{f}=3$ (Fig. \ref{fig:8}a) and the $n_{f}=4$ (Fig. \ref{fig:8}c). Thus, the stronger absorption region shifts to the smaller $b$, i.e. appear in more central collisions at $\sqrt{s}=14$ TeV with regard of the corresponding region at intermediate energy $\sqrt{s}=52.8$ GeV. Adopting the conservative estimation (\ref{eq:3b.3}), the behavior of the maximum of $G_{\scriptsize{\mbox{inel}}}^{\,\scriptsize{\mbox{c}}}(s,b)$ is
the same in dependence of the $\sqrt{s}$. Nonetheless, the excess of the inelastic overlap function at $\sqrt{s}=14$ TeV over the quantity at $\sqrt{s}=52.8$ GeV is seen in a significantly narrower region $0.07 \lesssim b \lesssim 0.15$ fm. The relation is the opposite between these overlap functions for larger $b$ and near the behavior of $G_{\scriptsize{\mbox{inel}}}^{\,\scriptsize{\mbox{c}}}(s,b)|_{\sqrt{s}=52.8\,\scriptsize{\mbox{GeV}}}$ and
$G_{\scriptsize{\mbox{inel}}}^{\,\scriptsize{\mbox{c}}}(s,b)|_{\sqrt{s}=14\,\scriptsize{\mbox{TeV}}}$
for smaller $b$. These statements are valid at $n_{f}=3$ (Fig. \ref{fig:8}b) and at $n_{f}=4$ (Fig. \ref{fig:8}). Thus, the conservative scheme for the $\omega$ lead to the hollowness effect for both very different energies considered here.

\begin{figure}[ht]
\centering
\includegraphics[width=14.0cm,height=12.0cm]{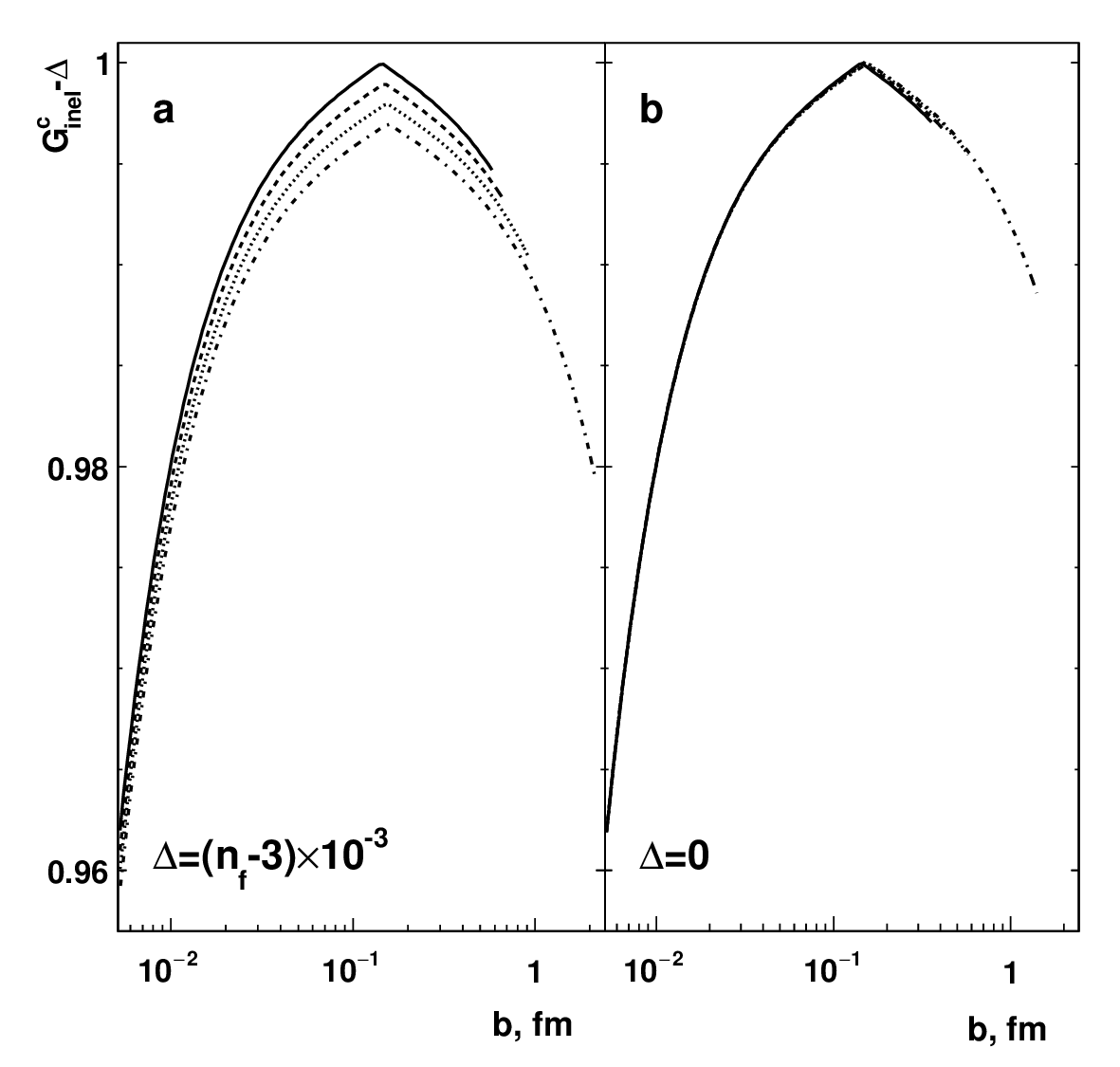}
\caption{\label{fig:after8}Dependence of
$G_{\scriptsize{\mbox{inel}}}^{\,\scriptsize{\mbox{c}}}$ on $b$
for the confinement potential at $b_{\scriptsize{\mbox{min}}}=2.0
\times 10^{-4}$ fm and $\sqrt{s}=14$ TeV for various $n_{f}$ with
$\omega=1$ (a) and conservative estimation (\ref{eq:3b.3}) used
for $\mu_{\scriptsize{\mbox{min}}}$ (b). The solid lines are for
$n_{f}=3$, dashed ones -- for $n_{f}=4$, dotted curves correspond
to the $n_{f}=5$ and dot-dashed ones -- to the $n_{f}=6$. The
curves are shifted on the finite $\Delta$ in (a) for clearness.}
\end{figure}

All curves shown in the Figs. \ref{fig:7} and \ref{fig:8} reveal
the presence of a critical value $b_{0}$, in agreement with the
analyses performed, revealing the existence of a gray area for $b
<b_{0}$. As above mentioned, the
$b_{\scriptsize{\mbox{min}}}=0.05$ fm is mostly chosen for the
display of $n_{f}$-- and $s$--dependencies for
$G_{\scriptsize{\mbox{inel}}}^{\,\scriptsize{\mbox{c}}}$. As
expected from Fig. \ref{fig:after6}, the view of $b$--dependence
of the inelastic overlap function for the confinement potential
changes with $b_{\scriptsize{\mbox{min}}}$ dramatically. Fig.
\ref{fig:after8} shows the
$G_{\scriptsize{\mbox{inel}}}^{\,\scriptsize{\mbox{c}}}(b)$ at
very small $b_{\scriptsize{\mbox{min}}}=2 \times 10^{-4}$ fm,
$\sqrt{s}=14$ TeV and several $n_{f}$ for $\omega=1$ (a) and
conservative estimation (\ref{eq:3b.3}) used for
$\mu_{\scriptsize{\mbox{min}}}$ (b). Accounting for the absence of
visible $n_{f}$--dependence in Fig. \ref{fig:after6}d, the curves
for various $n_{f}$ are shifted on finite $\Delta$ in Fig.
\ref{fig:after8}a. At present the $b_{\scriptsize{\mbox{min}}}=2
\times 10^{-4}$ fm can be considered as quite reasonable
approximation for $b=0$. In this case
$G_{\scriptsize{\mbox{inel}}}^{\,\scriptsize{\mbox{c}}}(b)$ shows
the approaching for the black disk limit at accuracy level
$\approx 2\%$ for $b$ varying in wide range $10^{-2}$ fm $\lesssim
b \leq b_{\scriptsize{\mbox{max}}}$. The range of $b$ in which
$G_{\scriptsize{\mbox{inel}}}^{\,\scriptsize{\mbox{c}}}(b) \approx
1$ is expanded significantly for $b_{\scriptsize{\mbox{min}}}=2
\times 10^{-4}$ fm in comparison with Figs. \ref{fig:7} and
\ref{fig:8}. However,
$G_{\scriptsize{\mbox{inel}}}^{\,\scriptsize{\mbox{c}}}(b)$ also
decreases sharply in the narrow region close to the
$b_{\scriptsize{\mbox{min}}}$ in this case (Fig.
\ref{fig:after8}).

Thus, the results shown in Figs. \ref{fig:7}--\ref{fig:after8} allows the following general conclusions. First of all, notice that inelastic overlap function description, based on the confinement potential, exhibits the black disk limit for $b\neq 0$ and a clear gray area emerging near $b=0$ (the hollowness effect). The gray area narrows with the decreasing of $b_{\scriptsize{\mbox{min}}}$ but it survives for any finite values of the parameter $b_{\scriptsize{\mbox{min}}}$. Then, the hollowness effect can be considered an essential and intrinsic feature of the confinement potential approach. It should be stressed that the potentials $V_{\scriptsize{\mbox{c}}}(s,b)$ and sequential $\tilde{V}_{\scriptsize{\mbox{c}}}(s,b)$ can be applied to describe the interaction of color-charged constituents, strictly speaking. At present, the self-consistent description from the first principles of QCD is absent for the transition from the quark-gluon plasma to the hadronic matter. Actually, there is an important hypothesis of local parton--hadron duality (LPHD) suggesting that the hadronization preserves the main features of the partonic interactions at hadronic level, i.e. the so-called soft hadronization \cite{ZPC-27-65-1985}. Nevertheless, the hadronization can influence and (slightly) distort a distribution for some quantities at experimentally measurable (hadronic) level, concerning the corresponding distributions for partonic interactions. Furthermore, one may suggest this influence can be amplified in some kinematic domain. Obviously, this is only a qualitative hypothesis which should be justified and verified by quantitative estimations. But, in any case, the comparison is possible at a qualitative level for $G_{\scriptsize{\mbox{inel}}}^{\,\scriptsize{\mbox{c}}}(s,b)$ obtained taking into account the confinement potential and the experimental results for hadronic (in particular, $pp$ and $\bar{p}p$) collisions. Considering the above and within the understanding that the use of confinement potential is a naive approach for $G_{\scriptsize{\mbox{inel}}}^{\,\scriptsize{\mbox{c}}}(s,b)$, one can make a qualitative comparison between the results of present work with some other phenomenological approaches.

In general, the behavior of $G_{\scriptsize{\mbox{inel}}}^{\,\scriptsize{\mbox{c}}}(s,b)$ in Fig. \ref{fig:7}, obtained taking into account the potential approach, confirms the results from \cite{dremin_1,dremin_2} and analyses done by \cite{broniowski_arriola,alkin_martynov,anisovich_nikonov,troshin_tyurin_1,anisovich,troshin_tyurin_2,albacete_sotoontoso,arriola_broniowski}. Furthermore, one expects that the inelastic overlap function description holds better for small values of $b$. In particular, the behavior of $G_{\scriptsize{\mbox{inel}}}^{\,\scriptsize{\mbox{c}}}(b)$ in Fig. \ref{fig:after8} shows the minimum at $b \to 0$ which is quite similar to the shallow minimum $G_{\scriptsize{\mbox{inel}}}^{\,\scriptsize{\mbox{c}}}(b) \simeq 0.97-0.98$ for most central bin $b \in [0.0;0.1]$ fm, obtained within mass squared approach with central optical potential at the same $\sqrt{s}=14$ TeV \cite{arriola_broniowski}. Also there is the second characteristic displayed by the hollowness effect in Fig. \ref{fig:after8}, namely the shift of the maximum to larger $b=b_{0}$ but $G_{\scriptsize{\mbox{inel}}}^{\,\scriptsize{\mbox{c}}}(b)$, obtained for hadronic level, shows a flatter growth and a wide maximum at $b \simeq 0.4-0.5$ fm \cite{arriola_broniowski}. Therefore, the confinement potential provides the hollowness effect on central collisions at the nominal LHC energy $\sqrt{s}=14$ TeV (Fig. \ref{fig:7}c, d), as obtained by another method for $pp$ collisions \cite{arriola_broniowski}. Furthermore, the analysis of high-statistic data close to $\sqrt{s}=13$ TeV and taking into account the L\'evy imaging \cite{T.Csorgo.R.Pasechnik.A.Ster.Euro.Phys.J.C80.126.2020}, results in a noticeably shallower minimum for $G_{\scriptsize{\mbox{inel}}}^{\,\scriptsize{\mbox{c}}}(0)=0.9915 \pm 0.0008$ at $b \leq 0.05$ fm and the maximum shifts to $b \simeq 0.4$ fm. But the last result does not contradict the prediction in Fig. \ref{fig:after8}, at qualitative level.

\begin{figure}[ht]
\centering
\includegraphics[width=14.0cm,height=12.0cm]{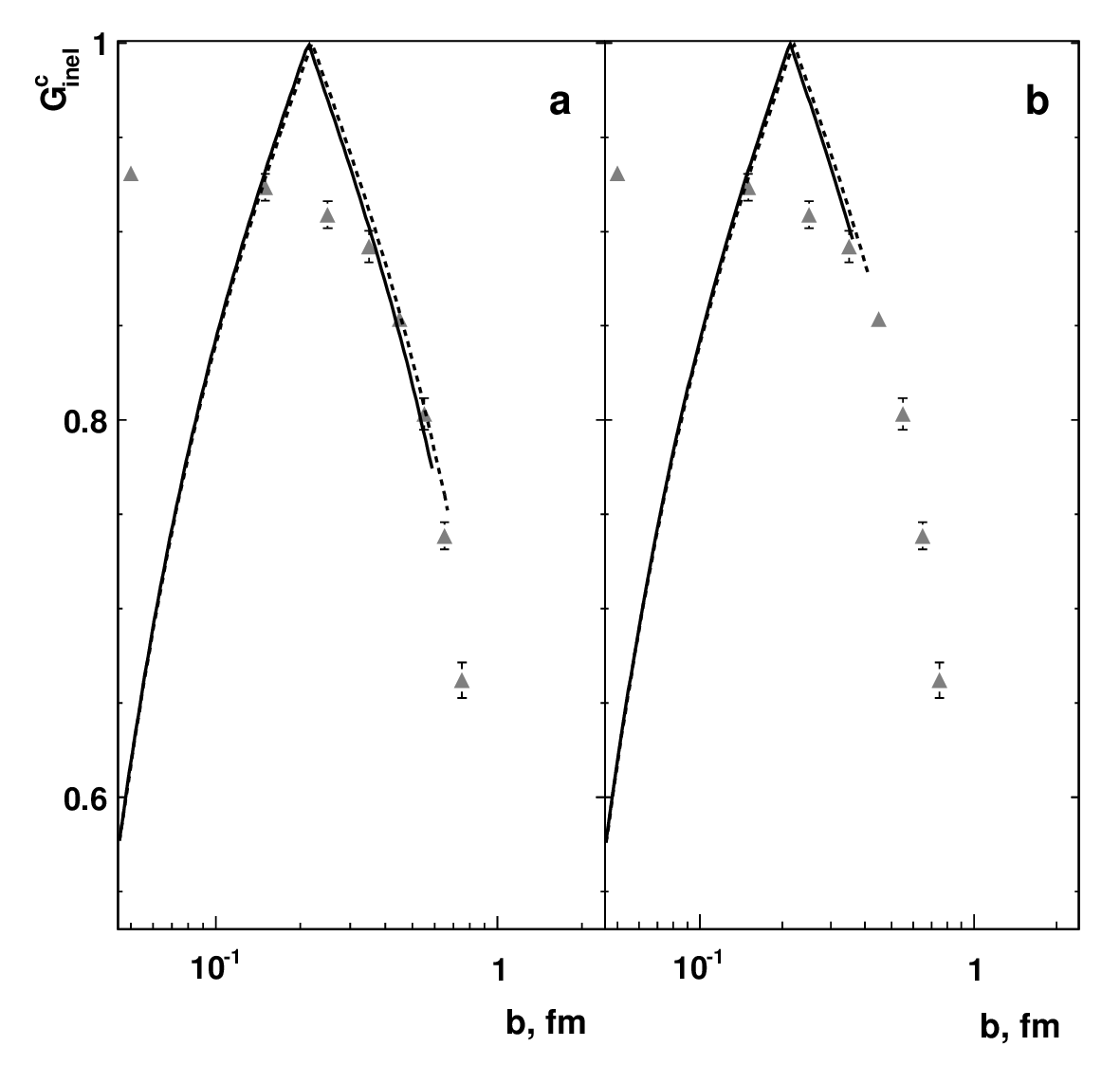}
\caption{\label{fig:after8-add1}Dependence of
$G_{\scriptsize{\mbox{inel}}}^{\,\scriptsize{\mbox{c}}}$ on $b$
for the confinement potential at $b_{\scriptsize{\mbox{min}}}=2.0
\times 10^{-2}$ fm and $\sqrt{s}=52.8$ GeV for various $n_{f}$
with $\omega=1$ (a) and conservative estimation (\ref{eq:3b.3})
used for $\mu_{\scriptsize{\mbox{min}}}$ (b) in comparison with
the results for $pp$ collisions at same $\sqrt{s}$ from
\cite{NPB-166-301-1980} shown by triangles. The solid lines are
for $n_{f}=3$, dashed ones -- for $n_{f}=4$.}
\end{figure}

\begin{figure}[ht]
\centering
\includegraphics[width=14.0cm,height=12.0cm]{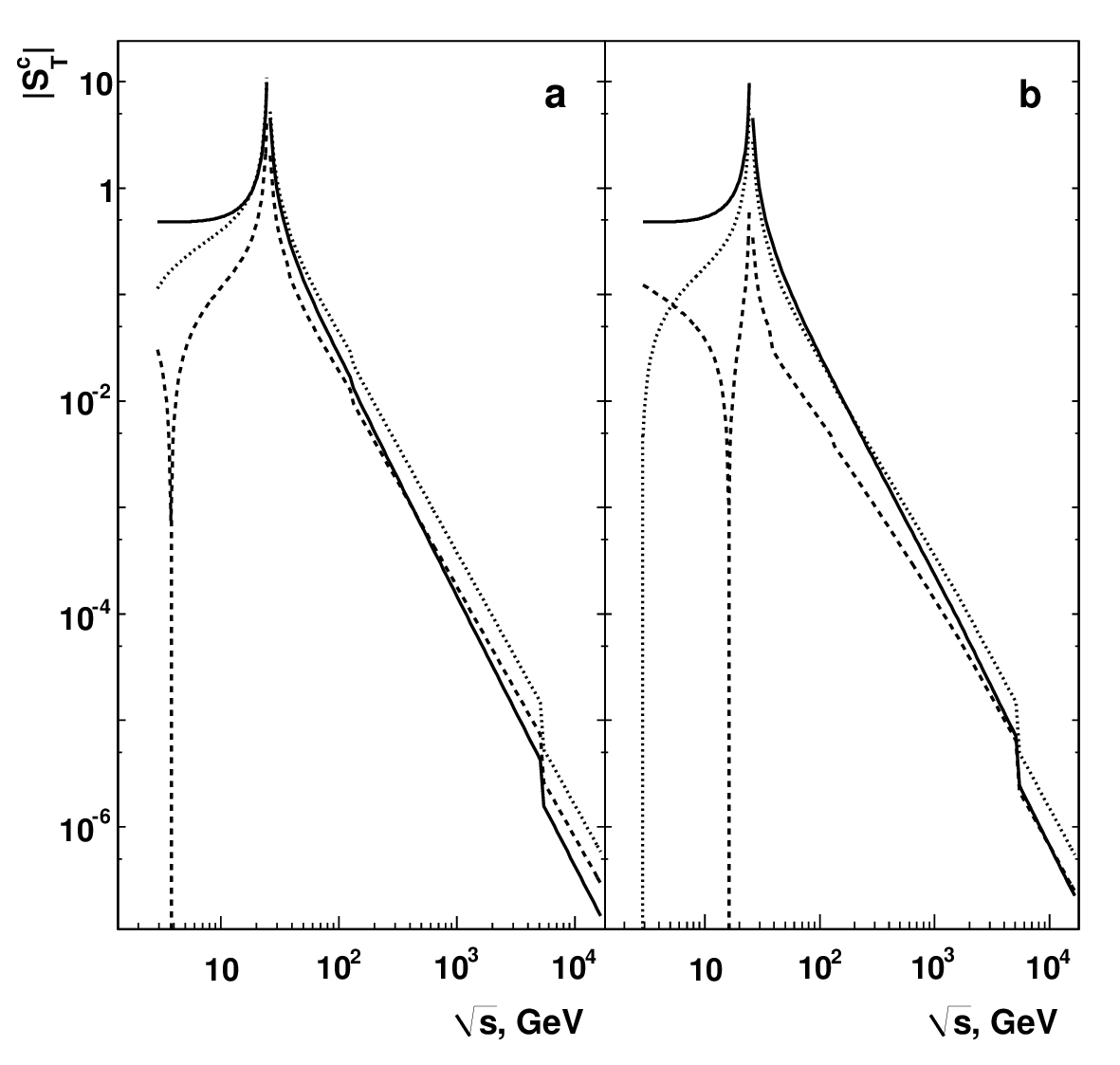}
\caption{\label{fig:9}Absolute values of the TE deduced for confinement potential in the impact parameter space on
collisions energy considering $b_{\scriptsize{\mbox{min}}}=0.05$ fm, $\sqrt{s_{c}}=25.0$ GeV, softest (a) and conservative (b) restriction on the $\mu_{\scriptsize{\mbox{min}}}$. The solid lines are for $b=0.10$ fm while dashed curve corresponds to the $b=0.35$ fm in (a) and $b=0.25$ fm in (b), dotted lines are for $b=0.60$ fm (a) and 0.40 fm (b).}
\end{figure}

It is known the shape of the $G_{\scriptsize{\mbox{inel}}}(b)$ is
model-dependent. Fig. \ref{fig:after8-add1} shows of
$G_{\scriptsize{\mbox{inel}}}^{\,\scriptsize{\mbox{c}}}$ on $b$
calculated within the present work for the confinement potential
at $b_{\scriptsize{\mbox{min}}}=2.0 \times 10^{-2}$ fm and
$\sqrt{s}=52.8$ GeV for various $n_{f}$ for soft (a) and
conservative (b) estimation used for
$\mu_{\scriptsize{\mbox{min}}}$ and results derived with help of
the another approach elsewhere \cite{NPB-166-301-1980}. As
discussed above the behavior of
$G_{\scriptsize{\mbox{inel}}}^{\,\scriptsize{\mbox{c}}}(b)$
depends on the free parameter $b_{\scriptsize{\mbox{min}}}$.
Therefore there is some arbitrariness at choosing a value of this
parameter for certain $\sqrt{s}$. This uncertainty can be
excluded, for instance, with help of the fit by (\ref{eq:3.2})
some reliable data. The value $b_{\scriptsize{\mbox{min}}}=2.0
\times 10^{-2}$ fm is chosen empirically because the definition of
the best value of $b_{\scriptsize{\mbox{min}}}$ for certain
$\sqrt{s}$ is outside the subject of the present work. The
approach for
$G_{\scriptsize{\mbox{inel}}}^{\,\scriptsize{\mbox{c}}}(b)$ based
on the perturbative confinement potential provides the
quantitative agreement with the results from
\cite{NPB-166-301-1980} at $b \gtrsim 0.3$ fm for $\omega=1$ (Fig.
\ref{fig:after8-add1}a). The indication on the similar conclusion
can be only suggested for Fig. \ref{fig:after8-add1}b because of
one point from \cite{NPB-166-301-1980} is in the region of overlap
of two models at $b \gtrsim 0.3$ fm. Predictions of two models
differ dramatically at smaller $b$: the present approach shows the
maximum for
$G_{\scriptsize{\mbox{inel}}}^{\,\scriptsize{\mbox{c}}}(b)$ at $b
\sim 0.2$ fm with subsequent decrease while the results from
\cite{NPB-166-301-1980} are smoothly increase. It means the
hollowness effect within the present approach based on the
$\tilde{V}_{\scriptsize{\mbox{c}}}(s,b)$ defined by
(\ref{eq:3b.7.new}) and absence of the effect for data points from
\cite{NPB-166-301-1980}. As qualitatively discussed above, the hadronization can influence on the view of $G_{\scriptsize{\mbox{inel}}}^{\,\scriptsize{\mbox{c}}}(b)$. Thus, the hadronization may explain, at least some part, the difference of the inelastic overlap function for quark and hadronic levels, in addition to the ambiguous choice of the $b_{\scriptsize{\mbox{min}}}$. It is interesting to note that the smooth approximations within the method from \cite{sdcvaocvm} demonstrate a noticeable deviation from the level $G_{\scriptsize{\mbox{inel}}}^{\scriptsize{\mbox{max}}}(s,0)=1$ in the energy domain from $\sqrt{s} \simeq 20$ GeV up to $\sqrt{s} \sim 100$ GeV for $pp$, $\bar{p}p$ and the combined sample for nucleon-nucleon scattering.

The $b$-dependence of the TE for confinement potential
($S_{T}^{\,\scriptsize{\mbox{c}}}$) is driven by the Figs.
\ref{fig:7}, \ref{fig:8} and relation (\ref{eq:3a.2}).

Likewise the Coulomb potential, the sign of the
$S_{T}^{\,\scriptsize{\mbox{c}}}$ is defined by $k$ due to the
normalization procedure. Furthermore, the TE for the confinement
potential is featured by sharply changes near $s_{c}$. The
absolute values of the TE for $s < s_{c}$
($|S_{T}^{\,\scriptsize{\mbox{c}}}|=-S_{T}^{\,\scriptsize{\mbox{c}}}$)
are larger by orders of magnitude than that for $s > s_{c}$
($|S_{T}^{\,\scriptsize{\mbox{c}}}|=S_{T}^{\,\scriptsize{\mbox{c}}}$).
Then, as well as in Sec. \ref{sec:4}, the
$|S_{T}^{\,\scriptsize{\mbox{c}}}|$ is the adequate quantity for
study of $s$-dependence of the TE for the approximation
(\ref{eq:3b.4}) of the confinement potential in $b$-space.

The Fig. \ref{fig:9} shows the energy behavior of the TE absolute
value for the confinement potential within the 5-loop
approximation for $\alpha_{s}$ at
$b_{\scriptsize{\mbox{min}}}=0.05$ fm, $\sqrt{s_{c}}=25.0$ GeV,
softest (a) and conservative (b) restriction on the
$\mu_{\scriptsize{\mbox{min}}}$ for several $b$. The additional
analysis shows that $|S_{T}^{\,\scriptsize{\mbox{c}}}|(s,b)$ does
not depend on the scheme for the estimation of $\omega$ at a given
value of the impact parameter\footnote{See, for instance, the
curves at $b=0.10$ fm in Fig. \ref{fig:9}a and \ref{fig:9}b.}.
Thus, the values of $b$ differ in Fig. \ref{fig:9}a and
\ref{fig:9}b for most cases. As expected, the factor $k$ provides
similar general trends for the energy dependence of the TE in Fig.
\ref{fig:9} in comparison with Fig. \ref{fig:3}. However, the
behavior of $|S_{T}^{\,\scriptsize{\mbox{c}}}(s,b)|$ is more
intricate than that for the Coulomb potential. The very sharp
minimums are the $|S_{T}^{\,\scriptsize{\mbox{c}}}(s,b)|=0$ at
$\sqrt{s}$ for which $b=b_{0}$. As discussed above, the sharp
changing of the $|S_{T}^{\,\scriptsize{\mbox{c}}}(s,b)|$ due to
onset of influence of heavier flavors is most visible for
$n_{f}=6$ in TeV-energy range.
\\
\\
\section{\label{re}Discussions and Conclusions}

The presence of the hollowness effect (gray area) cannot be
associated with limiting the resolution of the facilities. On the
other hand, the de Broglie wavelength achieves its minimum at
present-day energies at the LHC, and despite its small value
$\Delta r=1/p_{cm} \sim 2/\sqrt{s}$ it still produces an
unavoidable natural coarse-grain effect.

The use of potentials mimicking the internal energy is not new in
physics, probably remounting to Bohm quantum potential
\cite{bohm_1,G.Dennis.M.A.de.Gosson.B.J.Hiley.Phys.Lett.A378.2363.2014}
in Quantum Mechanics and, more specifically, in nuclear scattering
\cite{Shuryak-book-2004-p374}. However, the use of both the
Coulomb and the confinement potentials are illustrative of the
physical behavior of $pp$ and $\bar{p}p$ in the impact parameter
space. It should be stressed that the confinement potential was applied here at the level of the charged-color particles. Moreover, the hadronization can influence and distort distributions at the hadronic level despite the LPHD hypothesis. Therefore the results obtained in the resent work are not
able to fit the experimental data for the inelastic overlap
function since this is not the aim of a potential approach.

The Coulomb potential treats the hadrons as point-like objects,
and its description using the impact parameter picture does not
allow any acceptable result near the forward direction. Far from
the forward direction  (high $q^2$), derivative terms can also be
added to achieve a better description of the inelastic overlap
function introducing a slowdown decaying for the tail (low $q^2$).

On the other hand, the confinement potential considered here as
the internal energy of the hadron shows the hollowness effect in a qualitative level.
Thus, the confinement potential may allow the rise of this
effect even if we add derivative terms (corresponding to the tail). It is important to stress
that, using a different approaches, the hollowness effect was
recently predicted at the LHC energies $\sqrt{s}=7$ \cite{arriola_broniowski} 13
\cite{PRD-98-074012-2018,T.Csorgo.R.Pasechnik.A.Ster.Euro.Phys.J.C80.126.2020} and 14 TeV \cite{arriola_broniowski}.

The physical behavior expressed by the confinement potential represents an intrinsic feature of the strong interaction. Then, the confinement potential approach furnishes the {\it qualitative general behavior} of the inelastic overlap function. The results shown in Figs. \ref{fig:7} -- \ref{fig:after8} represent the impossibility to ascribe to the inelastic overlap function only one exponential \cite{sdcvaocvm,D.A.Fagundes.M.J.Menon.P.V.R.G.Silva.Nucl.Phys.A946.194.2016}. The presence of a persistent maximum even in the energy region where it is not expected can be attributed to the inevitable normalization procedure (see Fig. \ref{fig:after8-add1}). Then, one expects the cuspid behavior can be smoothed for low energies. Otherwise, the cuspid becomes pronounced as the energy rise. Note this behavior is absent in the Coulomb approach.

In order to avoid the hollowness effect, from the confinement
point of view, we should modify the confinement potential adding
correction terms acting only near the forward direction. These
terms may correspond, for example, to kinematic terms emerging at
a very high $q^2$ (very short distance). However, it seems
unlikely since corrections to the linear term of (\ref{eq:3b.4})
imply or in the decreasing of the strength of the confinement
potential to $b\rightarrow 0$ or simply not modifying its general
behavior as $b\rightarrow 0$ (or introducing some noise or small
perturbations). None of these assumptions seems to be physically
reasonable. Therefore, we claim here that the presence of a gray
area in the impact parameter space is a consequence of the
thermodynamic processes as well as of the multifractal character
of the hadron in the energy and momentum spaces.

The entropy probably is one of the most important physical
quantities in nature and should be taken into account in all
physics explanations. In the TE, the entropic index $w$ is
replaced by a convenient choice of parameters representing a phase
transition occurring at $s=s_c$, in total cross-section
experimental data-set. The probability density function is
replaced by the inelastic overlap function in the impact parameter
space. This convenient form of entropy provides an understanding
of how the matter density induces the geometric pattern observed
in the $pp$ and $\bar{p}p$ elastic scattering.

The increasing or decreasing entropy implies in an increasing or
decreasing probability of the inelastic overlap function, which
result is the emergence of a critical value $b_{0}$ associated
with the matter distribution inside $pp$ and $\bar{p}p$ elastic
scattering. Therefore, the entropy determines the existence of the
critical value in the impact parameter space. The consequence of
this result may be viewed as the presence of a fractal character
in the momentum space
\cite{antoniou_1,antoniou_2,antoniou_4,antoniou_5,bialas_1,bialas_2}.

Of course, the TE is one of several one ways to compute the
entropy of a non-additive system. However, without loss of
generality, the cases of interest can be reduced to the Tsallis
form, even the additive entropies by taking $w=1$
\cite{beck_0902.1235v2, tsallis_book}.

The $\bar{q}q$-interaction entails the energy density distribution
inside the proton and may determine the emergence of the
hollowness effect. Recently, the $k$-factor was introduced to take
into account the phase transition occurring in the total cross
section furnishing an explanation for the radial pressure
distribution in the proton
\cite{S.D.Campos.Int.J.Mod.Phys.A34.1950057.2019}. A possible
consequence that result is the emergence of the hollowness effect
manifested in the von Laue stability condition
\cite{M.von.Laue.Annalen.der.Physik.340.(8).524.1911}.

Finally, the analyses carried out here are based on few physical
assumptions and allows one to obtain the occurrence of the gray
area in the inelastic overlap function without the use of models
for the $pp$ and $\bar{p}p$ elastic scattering. It should be
emphasized that the approach presented is not able to furnish any
\textit{best fitting result} of any experimental data, since this
is not its aim, which is the qualitative study of both the
possible phase transition in the total cross section and the
existence of a gray area in the inelastic overlap function.
Bearing this in mind, the results obtained can help in the
construction of models taking into account the existence of both
physical phenomena in the $pp$ and $\bar{p}p$ elastic scattering.

\section*{Acknowledgments}

S.D.C. thanks to UFSCar by the financial support. The work of V.A.O. was supported partly by NRNU MEPhI Academic Excellence Project (contract No 02.a03.21.0005, 27.08.2013).


\begin{thebibliography}{0} 
\bibitem{CERN-YR-2019-007}{\it Report on the
physics at the HL--LHC, and perspectives for the HE--LHC}. Eds. A.
Dainese, M. Mangano, A. B. Meyer et al. CERN Yellow report:
monographs. CERN--2019--007. CERN, Geneva (2019).

\bibitem{anisovich_nikonov}V. V. Anisovich, V. A. Nikonov and J. Nyiri, Phys. Rev. D\textbf{90}, 074005 (2014).

\bibitem{V.V.Anisovich.M.A.Matveev.V.A.Nikonov.Int.J.Mod.Phys.A31.1645019.2016}V. V. Anisovich, M. A. Matveev and V.A. Nikonov, Int. J. Mod. Phys. A\textbf{31}, 1645019 (2016); G. Pancheri and Y. N. Srivastava, Eur. Phys. J. C\textbf{77}, 3 (2017).

\bibitem{sdcvaocvm}S. D. Campos, V. A. Okorokov and C. V. Moraes, Phys. Scr. \textbf{95}, 025301 (2020). 

\bibitem{Cheng-book-1987} H. Cheng and T. T. Wu, {\it Expanding
protons: scattering at high energies}. The MIT Press (1987); A.
Donnachie, H. G. Dosch, P. V. Landshoff and O. Nachmann, {\it
Pomeron physics and QCD}. Cambridge Univ. Press (2002).

\bibitem{Barone-book-2002}V. Barone and E. Predazzi, {\it High-energy particle diffraction}.
Springer--Verlag (2002).

\bibitem{dremin_1}I. M. Dremin, Phys. Uspekhi \textbf{58}, 61 (2015).

\bibitem{dremin_2}I. M. Dremin Phys. Uspekhi \textbf{60}, 333 (2017).

\bibitem{broniowski_arriola}W. Broniowski and E. Ruiz Arriola, Acta Phys. Polon. B\textbf{10} Proc. Supp., 1203 (2017).

\bibitem{alkin_martynov}A. Alkin, E. Martinov, O. Kovalenko and S. M. Troshin, Phys. Rev. D\textbf{89}, 091501 (2014).

\bibitem{troshin_tyurin_1}S. M. Troshin and N. E. Tyurin, Int. J. Mod. Phys. A\textbf{29}, 1450151 (2014).

\bibitem{anisovich}V. V. Anisovich, Phys. Uspekhi \textbf{58}, 963 (2015).

\bibitem{troshin_tyurin_2}S. M. Troshin and N. E. Tyurin, Mod. Phys. Lett. A\textbf{31}, 1650079 (2016).

\bibitem{albacete_sotoontoso}J. L. Albacete and A. Soto-Ontoso, Phys. Lett. B\textbf{770}, 149 (2017).

\bibitem{arriola_broniowski}E. Ruiz Arriola and W. Broniowski, Phys. Rev. D\textbf{95}, 074030 (2017).

\bibitem{bc}F. S. Borcsik and S. D. Campos, Mod. Phys. Lett. A\textbf{31}, 1650066 (2016).

\bibitem{tsallis_multifractal}C. Tsallis, Braz. J. Phys. \textbf{39}, 337 (2009).

\bibitem{antoniou_1} N. G. Antoniou, F. Diakonos and C. G. Papadopoulos, Phys. Lett. B\textbf{265}, 399 (1991).

\bibitem{antoniou_2}N. G. Antoniou, V. E. Zambetakis, F. K. Diakonos, and N. K. Diakonou, Z. Phys. C\textbf{55}, 631 (1992).

\bibitem{antoniou_4}N. G. Antoniou, F. Diakonos, I. S. Mistakidis and C. G. Papadopoulos, Phys. Rev. D\textbf{49}, 5789 (1994).

\bibitem{antoniou_5}N. G. Antoniou, N. Davis, and F. K. Diakonos, Phys. Rev. C\textbf{93}, 014908 (2015).

\bibitem{bialas_1}A. Bialas, Nucl. Phys. A\textbf{545}, 285c (1992).

\bibitem{bialas_2}A. Bialas, Acta Phys. Pol. B\textbf{23}, 561 (1992).

\bibitem{deppman_phys_rev_d93_054001_2016}A. Deppman, Phys. Rev. D\textbf{93}, 054001 (2016).

\bibitem{S.D.Campos.Phys.Scr.95.065302.2020}S. D. Campos, Phys. Scr. \textbf{95}, 065302 (2020).

\bibitem{S.D.Campos.arXiv.2003.11493.2020}S. D. Campos, arXiv: 2003.11493 [hep-ph] (2020).

\bibitem{froissart_1961} M. Froissart, Phys. Rev. \textbf{123}, 1053 (1961).

\bibitem{lukaszuk_1967}L. Lukaszuk and A. Martin, Nuovo Cim. A\textbf{52}, 122 (1967).

\bibitem{martin_2009}A. Martin, Phys. Rev. D\textbf{80}, 065013 (2009).

\bibitem{wu_2011}T. T. Wu, A. Martin, S. M. Roy, and V. Singh, Phys. Rev. D\textbf{84}, 025012 (2011).

\bibitem{PRD-91-076006-2015}A. Martin and S. M. Roy, Phys. Rev. D\textbf{91}, 076006 (2015).

\bibitem{arXiv-1805.10514} V. A. Okorokov, Phys. At. Nucl. \textbf{82}, 134 (2019).

\bibitem{NPB-166-301-1980}U. Amaldi and K. R. Schubert, Nucl. Phys. B\textbf{166}, 301 (1980).

\bibitem{PLB-132-443-1983}R. Henzi and P. Valin, Phys. Lett.
B\textbf{132}, 443 (1983).

\bibitem{PLB-160-167-1985}R. Henzi and P. Valin, Phys. Lett.
B\textbf{160}, 167 (1985).

\bibitem{PRD-89-051901-2014}A. Alkin, E. Martynov, O. Kovalenko and S. M. Troshin, Phys. Rev. D\textbf{89}, 051901 (2014).

\bibitem{PRD-98-074012-2018}W. Broniowski, L. Jenkovszky, E. Ruiz Arriola and I. Szanyi, Phys. Rev. D\textbf{98}, 074012 (2018).

\bibitem{T.Csorgo.R.Pasechnik.A.Ster.Euro.Phys.J.C80.126.2020} T. Cs\"org\H{o}, R. Pasechnik and A. Ster, Eur. Phys. J. C\textbf{80}, 126 (2020).

\bibitem{EPJC-79-861-2019}G. Antchev et al. (TOTEM Collaboration), Eur. Phys. J. C\textbf{79}, 861
(2019).

\bibitem{EPJC-78-913-2018}I. M. Dremin and V. A. Nechitailo, Eur. Phys. J. C\textbf{78}, 913 (2018).

\bibitem{renyi_entropy}A. R\'enyi, in {\it Proceedings of the IV Berkeley Symposium on mathematical statistics and probability}, \textbf{1}, 547 (1960).

\bibitem{shannon_entropy}C. E. Shannon, Bell Sys. Tech. J. \textbf{27}, 379 (1948).

\bibitem{abe_entropy}S. Abe, Phys. Lett. A\textbf{224}, 326 (1997).

\bibitem{beck_0902.1235v2}C. Beck, Contemporary Phys. \textbf{50}, 495 (2009). 

\bibitem{tsallis_book}C. Tsallis, {\it Introduction to nonextensive statistical mechanics: approaching a complex world}. Springer Science (2009).

\bibitem{PA-261-534-1998}C. Tsallis, R. S. Mendes and A.R. Plastino, Phys. A\textbf{261}, 534 (1998).

\bibitem{Levich-book-1971}B. G. Levich, {\it Theoretical physics: an advanced text}. \textbf{2}, John Wiley \& Sons, Inc. (1971).

\bibitem{JIKapusta-GCharles-book-2006}J. I. Kapusta and G. Charles, {\it Finite-temperature field theory principles and
applications}. Cambridge Univ. Press (2006).

\bibitem{kaufman_book_2001}M. Kaufman, {\it Principles of Thermodynamics}. Marcel Dekker, Inc. (2001).

\bibitem{Shuryak-book-2004-p374}E. V. Shuryak, {\it The QCD vacuum, hadrons and superdense matter}. Lec. Notes Phys. \textbf{71}, World Scientific (2004) and references therein.

\bibitem{PRD-70-054507-2004}E. V. Shuryak and I. Zahed, Phys. Rev. D\textbf{70}, 054507 (2004).

\bibitem{AIPCP-602-323-2001}F. Karsch, AIP Conf. Proc. \textbf{602}, 323 (2001).

\bibitem{PTPSuppl-153-287-2004} O. Kaczmarek et al., Prog. Theor. Phys. Supp. \textbf{153}, 287 (2004); O. Kaczmarek and F. Zantow, PoS (LAT2005), 192 (2005); Y. Burnier, O. Kaczmarek and A. Rothkopf, Phys. Rev. Lett. \textbf{114}, 082001 (2015); P. Petreczky, A. Rothkopf and J. Weber, Nucl. Phys. A\textbf{982}, 735 (2019).

\bibitem{NPA-941-179-2015} Shuai Y. F. Liu and R. Rapp, Nucl. Phys. A\textbf{941}, 179 (2015); Phys.
Rev. C\textbf{97}, 034918 (2018); Shuai Y. F. Liu, Min He and R.
Rapp, {\it ibid} \textbf{99}, 055201 (2019).

\bibitem{G.Dennis.M.A.de.Gosson.B.J.Hiley.Phys.Lett.A378.2363.2014}G. Dennis, M. A. de Gosson and B. J. Hiley, Phys.
Lett. A\textbf{378}, 2363 (2014); {\it ibid} \textbf{379}, 1224
(2015).

\bibitem{ramsey_phys_rev_103_10_1956}N. F. Ramsey, Phys. Rev. \textbf{103}, 10 (1956).

%

\bibitem{PRD-17-3090-1978}E. Eichten et al., Phys. Rev. D\textbf{17}, 3090 (1978).

\bibitem{PDG-PRD-98-030001-2018}
M. Tanabashi et al., Phys. Rev. D\textbf{98}, 030001 (2018).

\bibitem{PRD-90-074017-2014}A. P. Trawin$\acute{\mbox{n}}$ski et al., Phys. Rev. D\textbf{90}, 074017 (2014).

\bibitem{JHEP-1702-090-2017}F. Herzog, J. High Energy Phys. \textbf{02}, 090 (2017); P. A. Baikov, K. G. Chetyrkin and J. H. K$\ddot{\mbox{u}}$hn, Phys. Rev. Lett. \textbf{118}, 082002 (2017).

\bibitem{grunberg}G. Grunberg, Phys. Lett. \textbf{95}B, 70 (1980).

\bibitem{PRD-86-014022-2012}G. Aad et al. (ATLAS Collaboration), Phys. Rev. D\textbf{86}, 014022 (2012);
V. Khachatryan et al. (CMS Collaboration), Eur. Phys. J. C\textbf{75}, 288
(2015).

\bibitem{EPJC-75-186-2015}V. Khachatryan et al. (CMS Collaboration), Eur. Phys. J. C\textbf{75}, 186
(2015).

\bibitem{Leader-book-V2-1996}E. Leader and E. Predazzi E, {\it An introduction to gauge theories and modern particle physics}. \textbf{2}, Cambridge Univ. Press (1996).

\bibitem{J.Nyiri.Int.J.Mod.Phys.A.18.2403.2003}J. Nyiri, Int. J. Mod. Phys. A\textbf{18}, 2403 (2003).

\bibitem{EPJC-70-533-2010}E. K. G. Sarkisyan and A. S. Sakharov, Eur. Phys. J. C\textbf{70}, 533
(2010); E. K. G. Sarkisyan, A. N. Mishra, R. Sahoo, and A. S. Sakharov, Phys. Rev. D\textbf{93}, 054046 (2016).

\bibitem{M.Basile.etal.Lett. Nuovo.Cimen.38(10).359.1983}M. Basile et al., Lett. Nuovo Cimen. \textbf{38}, 359 (1983).

\bibitem{PAN-81-508-2018} V. A. Okorokov, Phys. At. Nucl. \textbf{81}, 508 (2018).

\bibitem{bartels_braun}J. Bartels and M. A. Braun, J. High Energy Phys. \textbf{06}, 095 (2018). 

\bibitem{V.L.Berezinskii.Zh.Eksp.Teor.Fiz.59.907.1970}V. L. Berezinskii, Sov. Phys. JETP \textbf{32}, 493 (1971). 

\bibitem{J.M.Kosterlitz.D.J.Thouless.J.Phys.C6.1181.1973}J. M. Kosterlitz and D. J. Thouless, J. Phys. C\textbf{6}, 1181 (1973).

%

\bibitem{D.A.Fagundes.M.J.Menon.P.V.R.G.Silva.Nucl.Phys.A946.194.2016}D. A. Fagundes, M. J. Menon and P. V. R. G. Silva, Nucl. Phys. A\textbf{946}, 194 (2016).

\bibitem{ZPC-27-65-1985}Ya. I. Azimov, Yu. L. Dokshitzer, V. A. Khoze and S. I. Troyan, Z. Phys. C\textbf{27}, 65 (1985).

\bibitem{bohm_1}D. Bohm, Phys. Rev. \textbf{85}, 166 (1952); {\it ibid}, 180 (1952).

\bibitem{S.D.Campos.Int.J.Mod.Phys.A34.1950057.2019}S. D. Campos, Int. J. Mod. Phys. A\textbf{34}, 1950057 (2019).

\bibitem{M.von.Laue.Annalen.der.Physik.340.(8).524.1911} M. von Laue, Ann. der Phys. \textbf{340}, 524 (1911).








\end{thebibliography}
\end{document}